\documentclass[%
aps,
prd,
amssymb,
amsmath,
amsfonts,
eqsecnum,
showpacs,
nofootinbib,
floatfix,
twocolumn,
twoside,
a4paper,
superscriptaddress,
longbibliography
]{revtex4-2}

\usepackage[T3,T1]{fontenc}
\usepackage{mathrsfs} 
\usepackage{bm} 
\makeatletter
\@ifclassloaded{beamer}
  { 
    \typeout{UsePackages: Detected beamer}
    \usepackage{tgheros}
    
  }
  { 
    \typeout{UsePackages: Did not detect beamer}
    \ifx\asybeamer\undefined 
    \typeout{UsePackages: Detected article}
    \usepackage[varg]{txfonts} 
    \usepackage{tgtermes} 
    \else 
    \typeout{Fonts: Detected asy for beamer}
    \usepackage{tgheros}
    
    \fi
  }
\usepackage{microtype} 
\def\MT@register@subst@font{
  \MT@exp@one@n\MT@in@clist\font@name\MT@font@list
  \ifMT@inlist@\else\xdef\MT@font@list{\MT@font@list\font@name,}\fi}
\makeatother

\usepackage{graphicx} %
\usepackage[x11names,svgnames,rgb]{xcolor} %
\usepackage{xspace}
\usepackage{braket}
\usepackage{accents}
\usepackage{siunitx}
\usepackage{mathtools}
\DeclareSymbolFontAlphabet{\mathrm}{operators}

\definecolor{CiteColor}{rgb}{0.18039, 0.18824, 0.57255}
\definecolor{UrlColor} {rgb}{0.741, 0.173, 0.000}
\definecolor{DarkUrlColor} {rgb}{0.500, 0.110, 0.000}
\definecolor{LinkColor}{rgb}{0.25098, 0.47843, 0.04706}

\makeatletter %
\newcommand{\ShowFont}{%
  \typeout{The main font is \f@encoding \space \f@family \space %
    \f@series \space \f@shape \space at \f@size pt.}%
  \typeout{The math font sizes are \tf@size pt (main), \sf@size pt %
    (script), and \ssf@size pt (scriptscript).}%
  \typeout{The linewidth is \the\linewidth}} %
\makeatother %

\usepackage{xargs}

\usepackage{graphicx}
\usepackage{dcolumn}
\usepackage{bm}
\usepackage{tabularx}
\usepackage{multirow}
\usepackage{capt-of}
\usepackage{color}
\usepackage{url}
\usepackage[utf8]{inputenc}
\usepackage{placeins}

\usepackage{ulem}

\usepackage{graphicx}
\usepackage{booktabs}
\usepackage{adjustbox}
\usepackage{amsmath}

\usepackage{natbib}

\usepackage[nolist,nohyperlinks]{acronym}
\usepackage{xspace}
\usepackage[colorlinks]{hyperref}

\usepackage[caption=false]{subfig}

\usepackage{float}

\DeclareMathAlphabet{\mathbfsf}{\encodingdefault}{\sfdefault}{bx}{sl}

\newcommand{\be}{\begin{equation}}
\newcommand{\ee}{\end{equation}}
\newcommand{\bea}{\begin{eqnarray}}
\newcommand{\eea}{\end{eqnarray}}

\newcommand{\phT}{\textsc{IMRPhenomT}\xspace}
\newcommand{\phTE}{\textsc{IMRPhenomTEHM}\xspace}
\newcommand{\phTHM}{\textsc{IMRPhenomTHM}\xspace}
\newcommand{\phTPHM}{\textsc{IMRPhenomTPHM}\xspace}

\newcommand{\NRSur}{\textsc{NRSur7dq4}\xspace}

\newcommand{\chieff}{\chi_\mathrm{eff}}

\newcommand{\lvk}[1]{\textcolor{pink}{LVK: #1}}

\definecolor{dodgerblue}{HTML}{1E90FF}
\definecolor{viennared}{HTML}{DA0A14}
\definecolor{ctorange}{HTML}{FF6C0C}
\definecolor{granadagreen}{HTML}{078931}
\definecolor{wales}{HTML}{ff0038}
\definecolor{valenciacfred}{HTML}{ee3524}
\definecolor{barcelonafcgold}{HTML}{edbb00}
\definecolor{jam}{HTML}{A50B5E}
\definecolor{austriawien}{HTML}{441678}
\definecolor{navyblue}{HTML}{4169E1}

\AtBeginDocument{%
  \hypersetup{
    citecolor=dodgerblue,
    linkcolor=dodgerblue,   
    urlcolor=dodgerblue}}

\newcommand{\soft}[1]{\textsc{#1}}

\newcommand{\GWTCFOUR}{GWTC-4.0\xspace}

\newcommand{\BILBY}{\soft{Bilby}\xspace}

\newcommand{\LALSUITE}{\soft{LALSuite}\xspace}

\newcommand{\PESUMMARY}{\soft{PESummary}\xspace}
\newcommand{\PYTHON}{\soft{Python}\xspace}

\newcommand{\GWPY}{\soft{GWpy}\xspace}

\newcommand{\PEAUTOMATOR}{\soft{PEAutomator}\xspace}
\newcommand{\PHENOMXPY}{\soft{phenomxpy}\xspace}

\newcommand{\IMRPhenomXHM}{\soft{IMRPhenomXHM}\xspace}

\newcommand{\IMRPhenomXE}{\soft{IMRPhenomXE}\xspace}
\newcommand{\IMRPhenomXPHM}{\soft{IMRPhenomXPHM}\xspace}
\newcommand{\IMRPhenomXPHMST}{\soft{IMRPhenomXPHM\_SpinTaylor}\xspace}
\newcommand{\IMRPhenomXOFOURa}{\soft{IMRPhenomXO4a}\xspace}
\newcommand{\IMRPhenomXPNR}{\soft{IMRPhenomXPNR}\xspace}
\newcommand{\IMRPhenomT}{\soft{IMRPhenomT}\xspace}
\newcommand{\IMRPhenomTHM}{\soft{IMRPhenomTHM}\xspace}
\newcommand{\IMRPhenomTPHM}{\soft{IMRPhenomTPHM}\xspace}
\newcommand{\IMRPhenomTEHM}{\soft{IMRPhenomTEHM}\xspace}

\newcommand{\SEOBNRFIVEPHM}{\soft{SEOBNRv5PHM}\xspace}

\def\gw#1{gravitational wave#1}
\def\nr#1{numerical relativity
 (NR)#1\gdef\nr{NR}}
\def\bh#1{black-hole
 (BH)#1\gdef\bh{BH}}
\def\bbh#1{binary black hole#1
 (BBH#1)\gdef\bbh{BBH}}
\def\pn#1{post-Newtonian (PN)#1\gdef\pn{PN}}
\def\imr#1{inspiral-merger-ringdown (IMR)#1\gdef\imr{IMR}}
\def\eob#1{effective-one-body
 (EOB)#1\gdef\eob{EOB}}
\def\td#1{time-domain (TD)#1\gdef\td{TD}}
\def\fd#1{frequency-domain (FD)#1\gdef\fd{FD}}
\def\pe#1{parameter estimation (PE)#1\gdef\pe{PE}}
\def\lvk#1{LIGO-Virgo-KAGRA (LVK)#1\gdef\lvk{LVK}}

\newcommand{\UIB}{Departament de F\'isica, Universitat de les Illes Balears, IAC3 -- IEEC, Crta. Valldemossa km 7.5, E-07122 Palma, Spain}

\newcommand{\AEI}{Max Planck Institut für Gravitationsphysik (Albert Einstein Institut), Am M\"uhlenberg 1, Potsdam, Germany}

\newcommand{\ICE}
{Institut de Ci\`encies de l'Espai (ICE, CSIC), Campus UAB, Carrer de Can Magrans s/n, 08193 Cerdanyola del Vall\`es, Spain}

\allowdisplaybreaks[1]

\begin{document}

\title[ML]
{Parameter estimation for the \GWTCFOUR catalog with phenomenological waveform models that include orbital eccentricity and an updated description of spin precession}

\author{Yumeng Xu}
\affiliation{\UIB}

\author{Jorge Valencia} 
\affiliation{\UIB}

\author{Héctor Estellés Estrella}
\affiliation{\ICE}
\affiliation{\UIB}

\author{Antoni Ramos Buades}
\affiliation{\UIB}

\author{Sascha Husa}
\affiliation{\ICE}
\affiliation{\UIB}

\author{Maria Rosselló-Sastre}
\affiliation{\UIB}

\author{Joan Llobera Querol}
\affiliation{\UIB}

\author{Felip Ramis Vidal} 
\affiliation{\UIB}

\author{Maria de Lluc Planas Llompart}
\affiliation{\AEI}
\affiliation{\UIB}

\author{Marta Colleoni}
\affiliation{\UIB}

\author{Eleanor Hamilton}
\affiliation{\UIB}

\author{Arnau Montava Agudo}
\affiliation{\UIB}

\author{Jesús Yébana Carrilero}
\affiliation{\UIB}

\author{Anna Heffernan}
\affiliation{\UIB}

\date{\today}

\begin{abstract}
The \GWTCFOUR catalog
of transient gravitational wave signals describes observations made in the first part of the fourth observing run of the LIGO–Virgo–KAGRA (LVK) gravitational wave detector network. Here we extend the LVK's \GWTCFOUR analysis to elliptic orbits, and an improved description of
spin precession in the frequency domain.
For this study we use state‐of‐the‐art waveforms from the IMRPhenom family  (specifically XPNR, TPHM, and TEHM), and
we consider the 84 confidently detected events that are consistent with binary‐black‐hole mergers. 
We present an extended catalog of updated posterior samples, quantify how incorporation of these waveform effects alters inferred source properties relative to previous analyses, and discuss waveform systematics.
\end{abstract}


\maketitle

\section{Introduction}
\label{sec:Introduction}

The \lvk{} \cite{Aasi_2015,Acernese_2015,kagra_2021} has recently released \GWTCFOUR \cite{LIGOScientific:2025hdt,LIGOScientific:2025yae,LIGOScientific:2025slb,ligo_scientific_collaboration_and_virgo_2025_17014085,LIGO:2024kkz}, a catalog of detected transient (CBC) events in the first part of the fourth observing run (O4a) of the \lvk{} gravitational wave detector network,
covering 24 May 2023 to 16 January 2024. The catalog is accompanied by papers that update our knowledge on the population of compact binaries \cite{LIGOScientific:2025pvj}
and the cosmic expansion rate gravitational wave propagation \cite{LIGOScientific:2025jau}. Special events papers present a further discussion of the events GW230529\_181500 \cite{LIGOScientific:2024elc},
GW230814\_230901
\cite{LIGOScientific:2025cmm},
and GW231123\_135430
\cite{LIGOScientific:2025rsn}.

\GWTCFOUR contains Bayesian \pe{} results for 86 compact-binary coalescence candidates, which were detected at a false‐alarm‐rate $<1\,\mathrm{yr}^{-1}$). In the present paper we consider the 84 events that have been found consistent with BBH mergers 
\cite{LIGOScientific:2025hdt}.

\GWTCFOUR provides \pe{} results for all events (above a given threshold) for two waveform models, \IMRPhenomXPHMST \cite{Pratten:2020fqn,Garcia-Quiros:2020qpx,Pratten:2020ceb,Colleoni:2024knd} and \SEOBNRFIVEPHM\cite{Pompili_2023,Ramos-Buades:2023ehm,Estelles:2025zah}.
For events where the source parameters are contained in the calibration region of the \NRSur model \cite{Varma:2019csw}, such results are also included.
Furthermore, for some events with asymmetric masses or evidence
of precession, results with \IMRPhenomXOFOURa \cite{Hamilton:2021pkf,Thompson:2023ase,Ghosh:2023mhc} are also available, and for the most massive event GW231123\_135430
the time-domain phenomenological model \IMRPhenomTPHM \cite{Estelles:2020twz,Estelles:2021gvs} has also been used. 
For most events there is good agreement between waveform models, but not for all. Furthermore, 
none of these waveform models include eccentricity, and only \NRSur is fully calibrated to precessing numerical relativity waveforms, however the parameter space region in terms of mass, mass ratio and spin where the \NRSur model can be applied is more constrained than for other models. 
\IMRPhenomXOFOURa is calibrated to \nr{} waveforms, but this calibration is restricted to single spin systems and covers only the last cycles of the inspiral, the merger and ringdown.
In consequence, statements about systematic differences between waveform models for precessing systems are still difficult.
Understanding the more challenging regions is important for understanding the population of compact binaries, see e.g.
\cite{Dhani:2025xgt} for the effect of waveform systematics on inferring the Hubble constant.

Here we include results for three further waveform models from the IMRPhenom family
(i) \IMRPhenomXPNR \cite{Hamilton:2025xru} provides
an updated description of spin precession by
combining the accurate \pn{}-based inspiral description of 
\IMRPhenomXPHMST and the partial NR calibration of \IMRPhenomXOFOURa, 
(ii) \IMRPhenomTPHM, which is an extension of 
\IMRPhenomTHM to misaligned spins, and
(iii) \IMRPhenomTEHM \cite{Planas:2025feq}, which is an extension of 
\IMRPhenomTHM to aligned spin eccentric binaries, and allows us to extend the analysis to elliptic orbits. Both spin precession and eccentricity are indicators of dynamical formation channels, and improving methods to accurately measure these properties is essential for understanding the populations of compact binaries.

We carry out Bayesian parameter estimation with the same methods used in GWTC-4.0, see \cite{LIGOScientific:2025yae,Veitch:2014wba,Thrane_2019} for details. We assume in particular that the data consist of an astrophysical GW signal that can be described by our waveform models, superposed with colored Gaussian noise, with a power spectral density that can be assumed constant during the event.
We then compute the posterior
distribution from Bayes’ theorem, given a prior distribution and a likelihood function formulated in terms of 
the standard frequency-domain noise-weighted inner product.

The paper is organised as follows: In Sec.~\ref{sec:models}
we briefly describe the waveform models that we will use in this work. In our methods section, Sec.~\ref{sec:methods}, we then describe the observational data we use, our methods for Bayesian inference, and the automated workflow developed for this analysis.
In Sec.~\ref{sec:results:QC} we present our results for quasi-circular waveform models in a co-precessing reference frame - these results correspond to those presented in \GWTCFOUR, but add the \IMRPhenomXPNR and \IMRPhenomTPHM waveform models and provide further insight into waveform systematics in the quasi-circular sector.
We then present our analysis and results. In Sec.~\ref{sec:results_eccentricity} we then present our results with eccentric waveform models, and we
summarize our work and
conclude in Sec.~\ref{sec:conclusions}.

In addition to the discussion of our results, we provide parameter estimation posteriors and configurations used for \BILBY \cite{bilby} on the Zenodo platform \cite{uib_o4a_pe_catalog}.
The code used for parameter estimation automation are available on \cite{pe_automator}.

Throughout this paper, component masses are denoted by $m_i$. We define the mass ratios $q = m_2/m_1 \leq 1$ and $Q = m_1/m_2 \geq 1$, and the total mass is denoted by $M = m_1 + m_2$. These masses are to be understood to be defined in the detector frame, i.e. they are redshifted with respect to the masses in the source frame. In the discussions below we will alert the reader when referring to source frame masses. While masses in the source frame are generally  more relevant for answering astrophysical questions, 
detector frame masses are more relevant for understanding observational issues such as waveform systematics.
The $z$-component of the dimensionless spin magnitudes are 
denoted~$\chi_i=S_i^z/m_i^2$.
In order to characterise dominant spin effects it is common to decompose the two spin vectors into projections $\chi^\parallel_i$ parallel  and $\chi^\perp_i$perpendicular  to the orbital angular momentum $\mathbf{L}$ (which is in practice often approximated by the Newtonian angular momentum).
The dominant contribution to the orbital phasing can be parameterised by the effective spin  (or effective inspiral spin) $\chi_\mathrm{eff}$ ~\cite{Ajith:2011}
\begin{align} 
   \chi_\mathrm{eff} = {}& \frac{m_1 \chi_1^\parallel + m_2 \chi_2^\parallel}{m_1+m_2},
\end{align}
while the effective precession spin $\chi_\mathrm{p}$ ~\cite{Schmidt:2014iyl}
\begin{equation}
   \chi_\mathrm{p} =  \frac{1}{m_1^2 A_1} \max\left(A_1 S_1^\perp, A_2 S_2^\perp\right), 
\end{equation}
where $A_1 = 2+3m_2/(2m_1)$ and $A_2 = 2+3m_1/(2m_2)$ can be used to parameterise the dominant precession effects.

\section{Waveform models}\label{sec:models}

\subsection{\IMRPhenomXPNR}\label{sec:XPNR}

For our baseline analysis, we employ \IMRPhenomXPNR~\cite{Hamilton:2025xru}, a quasi-circular precessing \bbh{} model implemented in LALSuite \cite{lalsuite}, and reviewed by the \lvk{}. As other models from the IMRPhenomX* family, this model  is formulated in the frequency domain, which allows the most efficient likelihood evaluations - avoiding the Fourier transforms required for time domain models to evaluate the likelihood in the frequency domain, where the noise power spectral density is a function of frequency. Frequency domain models also avoid the need to make waveforms longer when one aims to include higher modes consistently at a given start frequency, see Eq.~(\ref{eq:eqFreq}) below.
\IMRPhenomXPNR is based upon \IMRPhenomXPHM~\cite{Pratten:2020ceb,Pratten:2021pro}, and is highly efficient, making it ideal to analyse a large catalog of events.
It includes the $(\ell,|m|)=\{(2,2),(2,1),(3,3),(3,2),(4,4)\}$ multipoles and uses the standard precession-modeling approach of tracking the precession of the binary and breaking the model down into the underlying co-precessing waveform and the precession dynamics, which inform the transformation between the co-precessing waveform and the fully precessing inertial-frame waveform \cite{PhysRevD.84.124011,PhysRevD.84.024046}.
\IMRPhenomXPNR employs the SpinTaylor evolution of the precession dynamics during inspiral~\cite{Colleoni:2024knd} and a phenomenological description of the merger-ringdown dynamics calibrated to \nr{} ~\cite{Thompson:2023ase, Hamilton:2021pkf}.
For the co-precessing waveform, we employ the aligned-spin model \IMRPhenomXHM~\cite{Garcia-Quiros:2020qpx}, with modifications to the dominant harmonic~\cite{Hamilton:2021pkf} in the late-inspiral, merger and ringdown by calibration to the same set of \nr{} simulations~\cite{Hamilton:2021pkf}. 
This is further combined with a model for the asymmetries present in the dominant multipole~\cite{Ghosh:2023mhc}.
Altogether, this results in the most comprehensive description of precessing systems available in semi-analytic models.

\subsection{\IMRPhenomT}\label{sec:T}

The \phT family of time-domain phenomenological models \cite{Estelles:2020osj,Estelles:2020twz,Estelles:2021gvs,Rossello-Sastre:2024zlr,Planas:2025feq} offers an alternative to the computationally efficient frequency-domain models. The flexibility of modeling effects in time-domain facilitates the inclusion of complicated new physical effects such as spin-precession \cite{Estelles:2021gvs}, orbital eccentricity \cite{Planas:2025feq} or GW memory \cite{Rossello-Sastre:2024zlr}.  Regarding the quasicircular (QC) models, we employ the nonprecessing-spin \phTHM model  \cite{Estelles:2020osj,Estelles:2020twz}, which has been calibrated to NR and test-particle waveforms and it includes $(l,|m|)=\{(2,2),(2,1),(3,3),(4,4),(5,5)\}$ multipoles, and the QC precessing-spin \phTPHM model \cite{Estelles:2021gvs}, which is based on the `twisting-up' approximation \cite{Schmidt:2014iyl} and it has the same mode content in the co-precessing frame as the \phTHM model. 

The \phTE model  \cite{Planas:2025feq} extends the QC aligned spin model \phTHM to describe binaries on elliptical orbits. 
It includes the multipoles $(\ell,|m|)=\{(2,2),(2,1),(3,3),(4,4),(5,5)\}$ and 3rd order \pn{} orbit-averaged dynamics~\cite{Henry:2023tka}, and waveform modes in an eccentricity expansion up to $\mathcal{O}(e^6)$. 

The \phTE model has been shown accurate against public \nr{} waveforms with initial eccentricities 0.3 with an unfaithfulness $<2\%$, and its applicability is conservatively limited to eccentricities below $e=0.4$ at an orbit-averaged (2,2)-mode frequency of 10 Hz \cite{Planas:2025feq} due to limited number of \nr{} waveforms. However, in recent studies \cite{Planas:2025jny,Planas:2025plq} it has been used for larger values of eccentricity up to 0.7 without evidencing unphysical behaviors in the waveform. For subsequent analyses in this paper, we adopt a threshold of $e\leq 0.5$ for standard runs, and extend it to $e\leq 0.65$ in specific cases.

The waveform models from the IMRPhenomT-family we use in this work are implemented in Python within the \PHENOMXPY package \cite{phenomxpy}.

\section{Methods}\label{sec:methods}

\subsection{Observational data}\label{sec:data}

The \gw{} strain data for the \pe{} are obtained from GWOSC \cite{LIGOScientific:2025snk} through the \GWPY package \cite{gwpy} with sample rate 4096 Hz. The duration of the downloaded strain data for each event is the duration of the analysis centered at the trigger time plus 16 second buffer before and after the analysis time, except for GW230608\_205047 (7s before the analysis period) and GW230518\_125908 (5s before the analysis period) due to the data availability.

Glitch mitigation as described in \cite{LIGO:2024kkz,LIGOScientific:2025slb} has been applied by the LVK to 16 of the strain data sets for events we will study, see 
Table 8 of \cite{LIGOScientific:2025slb}.
Furthermore, for a total of 19 of the events we study, the LVK analysis raised the start frequency for the likelihood evaluation from the default 20 Hz for at least one detector due to data quality issues, see again
Table 8 of \cite{LIGOScientific:2025slb}.
This includes 3 events for which glitch mitigation was also applied. Our analysis uses the same start frequencies for the likelihood evaluation as the LVK analysis, and we list the events with glitch mitigation or modified start frequency in Table \ref{table:DQ}.

\begin{table}[htb]
\caption{GW events with data quality issues used in this work. Column 2 indicates the detector set. Column 3 indicates whether glitch mitigation was applied. Columns 4–5 list the start frequencies used by the LVK, for analysing the data of each detector; the default value of 20\,Hz is indicated by ``--''. Our analysis uses the same start frequencies as the LVK. See our main text for details.
}
\label{table:DQ}
\begin{ruledtabular}
\begin{tabular}{lcccc}
Event & Detectors & Glitch mitigation & $f_\mathrm{start}^\mathrm{H1}$ [Hz] & $f_\mathrm{start}^\mathrm{L1}$ [Hz] \\
\hline
GW230606\_004305 & H1,L1 & \textbf{Yes} & -- & -- \\
GW230702\_185453 & H1,L1 & No           & 30 & -- \\
GW230707\_124047 & H1,L1 & \textbf{Yes} & -- & -- \\
GW230708\_053705 & H1,L1 & \textbf{Yes} & -- & -- \\
GW230709\_122727 & H1,L1 & \textbf{Yes} & 50 & -- \\
GW230729\_082317 & H1,L1 & No           & 50 & -- \\
GW230731\_215307 & H1,L1 & No           & 40 & -- \\
GW230803\_033412 & H1,L1 & No           & 30 & -- \\
GW230806\_204041 & H1,L1 & \textbf{Yes} & -- & -- \\
GW230819\_171910 & H1,L1 & \textbf{Yes} & -- & -- \\
GW230831\_015414 & H1,L1 & No           & -- & 22 \\
GW230914\_111401 & H1,L1 & \textbf{Yes} & -- & -- \\
GW230920\_071124 & H1,L1 & No           & 40 & -- \\
GW230924\_124453 & H1,L1 & \textbf{Yes} & -- & -- \\
GW231014\_040532 & H1,L1 & No           & 50 & -- \\
GW231018\_233037 & H1,L1 & No           & 30 & -- \\
GW231020\_142947 & H1,L1 & \textbf{Yes} & 45 & -- \\
GW231113\_122623 & H1,L1 & \textbf{Yes} & -- & -- \\
GW231114\_043211 & H1,L1 & \textbf{Yes} & -- & -- \\
GW231118\_005626 & H1,L1 & No           & 30 & -- \\
GW231118\_071402 & H1,L1 & No           & 50 & -- \\
GW231118\_090602 & H1,L1 & \textbf{Yes} & -- & -- \\
GW231123\_135430 & H1,L1 & \textbf{Yes} & -- & -- \\
GW231127\_165300 & H1,L1 & No           & 50 & -- \\
GW231129\_081745 & H1,L1 & \textbf{Yes} & 60 & -- \\
GW231206\_233134 & H1,L1 & No           & 40 & 30 \\
GW231221\_135041 & H1,L1 & \textbf{Yes} & -- & -- \\
GW231223\_032836 & H1,L1 & \textbf{Yes} & -- & -- \\
GW231223\_075055 & H1,L1 & No           & 40 & 30 \\
GW231223\_202619 & H1    & No           & 40 & -- \\
GW231224\_024321 & H1,L1 & No           & 40 & -- \\
GW240107\_013215 & H1,L1 & No           & 40 & -- \\
\end{tabular}
\end{ruledtabular}
\end{table}

\subsection{Inference methods}\label{sec:inferfence_methods}

Our inference methods follow those used by the LVK for the \GWTCFOUR catalog
\cite{LIGOScientific:2025hdt,LIGOScientific:2025yae,LIGOScientific:2025slb}.
In particular we also use the
dynesty ~\cite{dynesty} nested-sampling package as implemented within the 
\BILBY Bayesian inference infrastructure \cite{Ashton:2018jfp,Romero-Shaw:2020owr}. The version we are using is \texttt{v2.6.0},
which is the latest reviewed version contains a bug fix for the signal-to-noise ratio (SNR) normalization. 
Note that this version fixes a problem in the treatment of windowing of Fourier transforms, which had previously resulted in inflated values for signal-to-noise ratios and likelihoods, see \cite{LIGOScientific:2025yae} and \cite{Talbot:2025vth} for further details. 

Setting up a \pe{} analysis with \BILBY requires a configuration file, priors, the detector strain data as described in the previous section, the power spectral density (PSD) for each detector, and the calibration envelopes. Here we will describe how we setup each ingredient.
We take the \BILBY config extracted from the \pe{} samples \texttt{C00:IMRPhenomXPHM-SpinTaylor} downloaded from the Zenodo release \cite{ligo_scientific_collaboration_and_virgo_2025_17014085} as a baseline, and update the configuration accordingly for each of our waveform models. We use all the available subdominant spherical harmonic modes for each waveform model. 

We employ the default nested sampling method, i.e. we use the acceptance-walk method for the Markov Chain Monte Carlo (MCMC) evolution. For production runs we use an average number of accepted steps per MCMC chain of {\tt naccept=60} and number of live points {\tt nlive=1000} unless stated otherwise. Before starting our production runs we performed tests with a computationally cheaper setting of {\tt naccept=10} and number of live points {\tt nlive=300}. We notice that the posterior from the cheaper setting agrees pretty well with the production setting, while the error of Bayes factor is higher.
Two independent seeds are used for each run. 

By default, the minimum frequency for the likelihood evaluation in the frequency domain and the reference frequency for measuring time dependent quantities (in particular precessing spins, eccentricity, and the mean anomaly) are chosen as 20 Hz, as in the \GWTCFOUR setup. See table \ref{table:DQ} for cases where the minimal frequency for the likelihood calculation is not chosen as the default value of 20 Hz for data quality reasons.

The calibration envelope contain the 1-sigma systematic calibration error for the detector \cite{LIGOScientific:2025slb}. We use the enclosed calibration envelope in the \pe{} samples from the \GWTCFOUR release. Due to the definition difference in the input of \GWTCFOUR analysis and the calibration envelope stored in the \pe{} samples, we changed the \texttt{calibration-correction-type} from the default option \texttt{data} to \texttt{template}. They are related by the reciprocal function. The PSDs are also obtained from the \pe{} samples in \GWTCFOUR release.

For the \pe{} runs with precessing waveform models, specifically \IMRPhenomXPNR and \IMRPhenomTPHM, we use the same priors as the extracted bilby config. The prior is uniform in redshifted component masses, spin
magnitudes, merger time, and coalescence phase. It is isotropic in spin orientations, binary orientation, and sky location. The distance prior is chosen to be a uniform merger rate in comoving volume and time. 
For aligned-spin waveform models \IMRPhenomTHM, \IMRPhenomTEHM, we replace the spin components in the priors by Bilby's \texttt{AlignedSpin} function ranging from spin magnitude 0 to 0.99 
For the waveform models with eccentricity, we employ a uniform prior in eccentricity $e_{\rm{10\si{Hz}}}\in [0,0.65]$ and a uniform prior in mean anomaly $l_{\rm{10\si{Hz}}}\in [0,2\pi]$ with periodic boundary conditions at a reference redshifted frequency of $10\,\si{Hz}$ (For the sake of simplicity, we will refer to this quantity as just reference frequency).  In addition we also consider other reference frequencies for consistency checks as described in Sec.~\ref{sec:results_eccentricity}.

\subsection{Bayes Factor}\label{sec:bayes_factor}

Bayesian model comparison provides a well defined way to quantify the relative support that the data \(d\) give to two competing hypotheses \(\mathcal{H}_1\) and \(\mathcal{H}_2\).
The Bayes factor \(\mathcal{B}\) is defined as the ratio of the marginal likelihoods (evidences) of the two hypotheses,
\begin{equation}
\mathcal{B} \equiv \frac{p(d\!\mid\!\mathcal{H}_{1})}{p(d\!\mid\!\mathcal{H}_{2})}, \qquad
    \log\mathcal{B} = \log Z(\mathcal{H}_1) - \log Z(\mathcal{H}_2),
\end{equation}
where \(Z(\mathcal{H})\) denotes the Bayesian evidence for hypothesis $\mathcal{H}$.

Interpreting Bayes factors requires care because they depend on the prior choices for each hypothesis. Likelihood computation also varies with different data, sampler settings, etc., and furthermore our Bayesian analysis assumes stationary 
Gaussian noise.
In this work, we purposely avoid rigid thresholds for evidence derived from Bayes factor and instead report the full $\log_{10}\mathcal{B}$ and uncertainties, and in the mean time provide the case by case discussion.

\subsection{Automatization}\label{sec:automatization}

This work involved running hundreds of \pe{} jobs, which were run on a heterogeneous set of externally managed supercomputers targeting high performance computing applications in the national Spanish RES network.
To minimize manual intervention, we developed the automated workflow tool \PEAUTOMATOR{} \cite{pe_automator}. The framework consists of a \PYTHON package, project configuration files, and scripts for data acquisition. Using these scripts, we automatically download PE samples and strain data from GWOSC and Zenodo for \GWTCFOUR, and extract the corresponding \BILBY{} configuration files, PSDs, and calibration envelopes into a single standardized data directory, as described above.
The \PEAUTOMATOR{} package enables users to submit jobs for each event and waveform model to a supercomputing cluster with a single command. This command automatically handles \BILBY{} configuration generation, data packaging, job submission, and the creation of a GitLab issue for tracking. \PEAUTOMATOR{} is designed to be flexible, allowing users to run jobs across different clusters and allocations, thereby maximizing available computing resources.
In addition, a companion command-line tool running on a local server monitors job status. It updates GitLab issues with progress and error reports, and downloads results as jobs complete. For automated diagnostics, a scheduled command generates diagnostic plots and \PESUMMARY{}~\cite{Hoy:2020vys} pages at regular intervals, further reducing the need for manual oversight.

\section{Results for quasi-circular waveform models}\label{sec:results:QC}

We now discuss our results when using models that describe the same phenomena as in the \GWTCFOUR analysis, i.e. we restrict ourselves to use quasi-circular waveform models with misaligned spins. Comparing to  \GWTCFOUR, we include results from the latest frequency-domain phenomenological model \IMRPhenomXPNR and the time-domain phenomenological model \IMRPhenomTPHM. The aim is to better understand observed waveform systematics, and we will consider different metrics to measure deviations between IMRPhenomXPNR,  IMRPhenomTPHM, and the models used in \GWTCFOUR.

Most of the signals that we analyze are consistent with regions of the parameter space that are well modeled, and where indeed all waveform models that we use here or have been used for \cite{LIGOScientific:2025slb} agree well at current detector sensitivities.
In particular, already the LVK analysis finds that most binary black-hole events (81 of 84) are consistent with
mild mass ratios $1 \leq Q \leq 2.8$, mild spins, and detector frame (redshifted) masses  between $14 M_{\odot}\leq M \leq 200M_{\odot}$ \cite{LIGOScientific:2025slb}, where the merger-ringdown, that is particularly hard to model for precessing binaries, contributes only a small part of the SNR. Nevertheless, five events display redshifted total masses above $200M_{\odot}$ with greater than 90\% probability: GW230704\_212616, GW230922\_040658, GW231005\_021030, GW231028\_153006 and GW231123\_135430. For these high-mass events, we expect the signals to display mostly the merger-ringdown stages, where contributions from the higher-modes and differences in spin-precession modeling reduce the agreement of different models \cite{Estelles:2021jnz,Mehta:2021fgz}. In particular, GW231028\_153006 and GW231123\_135430 display significant waveform systematic effects in the \GWTCFOUR analyses and in our analysis.

Most of the events that we analyze here were found to be consistent with vanishing effective spin, $\chieff = 0$, although 16 of them show $\chieff > 0$ with greater than 90\% probability, and only three sources with $\chieff < 0$ with greater than 90\% probability \cite{LIGOScientific:2025slb}. Two events have significant positive $\chieff$ measurements, GW231028\_153006 and GW231118\_005626, both identified as showing potential waveform systematic discrepancies in \GWTCFOUR, and the former (a very high mass event) identified also by our analysis as showing significant waveform systematics.

In terms of signal-to-noise ratio, 6 events exceed a network matched-filter signal-to-noise ratio of 20 with greater that 90\% probability: GW230627\_015337, GW230814\_230901, GW231028\_153006, GW231123\_135430, GW231206\_233901 and GW231226\_101520, 4 of them displaying waveform systematic effects in the following analysis.

\subsection{Quantifying differences between posteriors}
To quantify the agreement between models, we compute a set of scalar diagnostics that measure the dissimilarity between posterior distributions for selected intrinsic parameters, namely the source-frame total mass $M$, the mass ratio $q$, and the effective spin parameters $\chi_{\rm eff}$ and $\chi_{\rm p}$. 

As a general, information–theoretic measure, we employ the Jensen–Shannon (JS) divergence \cite{JSdiv} between two probability density functions $P_A(x)$ and $P_B(x)$, defined as
\begin{equation}
    \mathrm{JS}(A,B) = \tfrac{1}{2} D_{\mathrm{KL}}\!\left(P_A \,\big\|\, \rm{mix(A,B)}\right) 
                    + \tfrac{1}{2} D_{\mathrm{KL}}\!\left(P_B \,\big\|\, \rm{mix(A,B)}\right),
\end{equation}
where $\rm{mix(A,B)} = (P_A + P_B)/2$ is the mixture distribution, and $D_{\mathrm{KL}}$ denotes the Kullback–Leibler (KL) divergence.
The JS divergence quantifies the information loss when two posterior distributions are approximated by their mean distribution, and ranges from 0 (identical distributions) to $\log 2$ (completely non-overlapping posteriors). This quantity (and the related KL divergence) is a standard metric for identifying systematic effects widely employed in gravitational-wave analyses \cite{Romero-Shaw:2020owr,GW190412}.

Complementary to the JS divergence, we define two additional metrics with  intuitive interpretation in terms of credible intervals. The normalized median shift,
\begin{equation}
    \Delta = \frac{|\mathrm{med}(A) - \mathrm{med}(B)|}{0.5\,[w_A + w_B]},
\end{equation}
measures the relative displacement of the two posterior medians in units of their average 90\% credible interval widths $w_A$ and $w_B$. 
A value of $\Delta = 0.3$ therefore indicates that the medians differ by about 30\% of a typical posterior width, while $\Delta \simeq 1$ corresponds to a full posterior-width shift.

The credible-interval overlap fraction,
\begin{equation}
    f_{\mathrm{overlap}} = 
    \frac{\max\!\left(0,\, \min(u_A,u_B) - \max(l_A,l_B)\right)}
         {\min\!\left(u_A-l_A,\, u_B-l_B\right)},
\end{equation}
is computed from the 90\% credible intervals $[l_i,u_i]$ of each posterior and quantifies how much the intervals overlap.
The complement, $1 - f_{\mathrm{overlap}}$, represents the fractional non-overlap between the two intervals.  
For example, $1 - f_{\mathrm{overlap}} = 0.2$ means that only 80\% of the narrower credible interval is shared between models, while values approaching unity indicate disjoint posteriors. Similar diagnostic quantities based on measuring the distance between median values or the overlap between posterior distributions are commonly employed for assessing systematic effects \cite{Akcay:2025rve,Ursell:2025ufb}.

\begin{figure*}[t!]
    \centering
    \includegraphics[width=\linewidth]{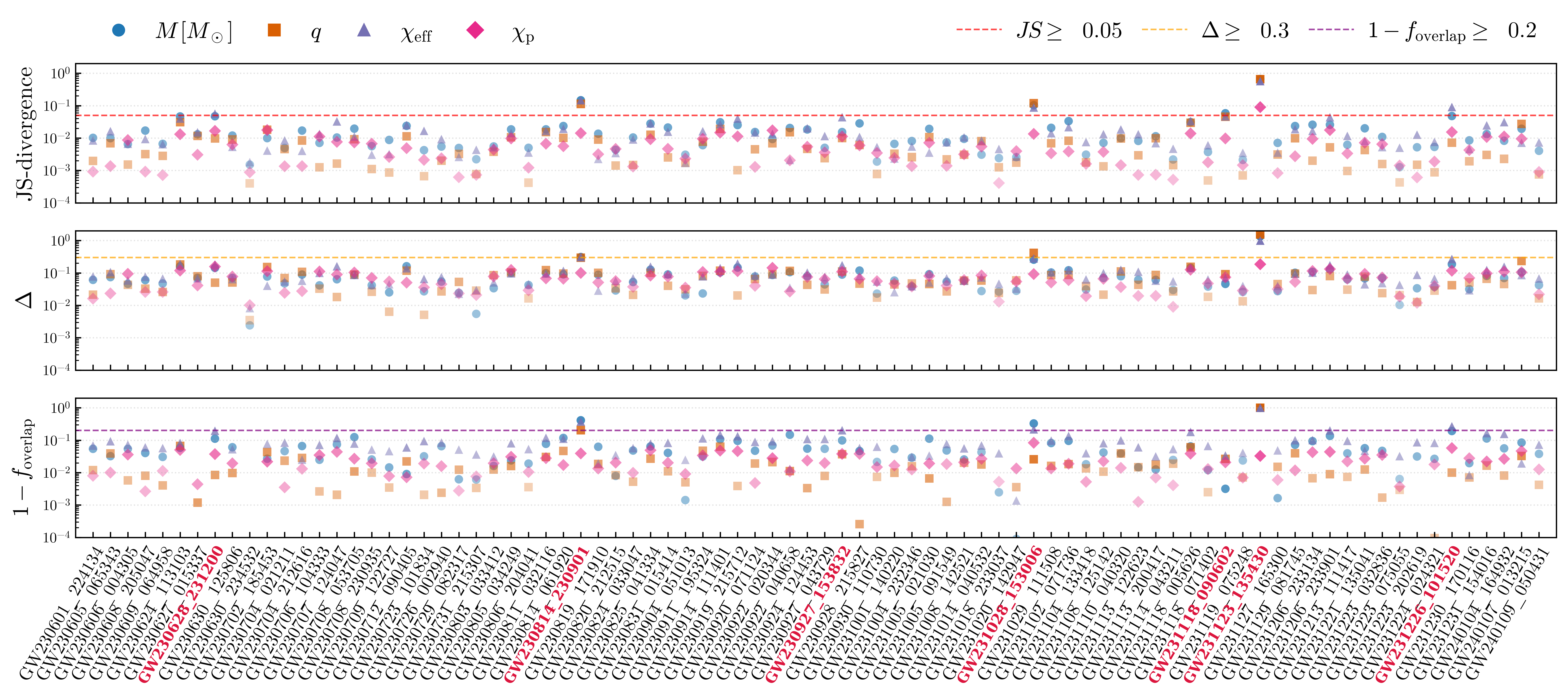}
    \caption{
        Scalar diagnostics used to quantify differences between posterior distributions from the \GWTCFOUR analyses and our new analyses with \IMRPhenomXPNR and \IMRPhenomTPHM.  
        Each panel shows, from top to bottom, the Jensen-Shannon divergence, the normalized median shift $\Delta$, and the complement of the credible-interval overlap fraction $1-f_{\rm overlap}$, evaluated for the source-frame total mass $M$, mass ratio $q$, and effective spin parameters $\chi_{\rm eff}$ and $\chi_{\rm p}$.  
        Horizontal dashed lines mark the adopted empirical thresholds 
        ($\mathrm{JS}\!\ge\!0.05$, $\Delta\!\ge\!0.3$, $1-f_{\rm overlap}\!\ge\!0.2$), 
        above which waveform-model differences are considered significant.  
        Points are colored by parameter, and events exceeding any threshold are highlighted in red along the $x$-axis labels.  
        The metrics collectively indicate that only a small subset of events exhibit measurable waveform-model systematics.  
    }
    \label{fig:metric_summary}
\end{figure*}

For each event, these three quantities are computed for all waveform-model pairs between \IMRPhenomXPNR, \IMRPhenomTPHM, and the models used in \GWTCFOUR \footnote{In this work we use the v2 \GWTCFOUR{} samples, i.e., the latest Zenodo release with the corrected likelihood normalization; see Appendix~\ref{app:likelhood} for further details.} 
for key intrinsic parameters: the redshifted total mass $M$, the mass-ratio $q$, and the effective spin parameters $\chi_{\rm eff}$ and $\chi_{\rm p}$. To be conservative, for each parameter and each diagnostic quantity we compute the maximum value across model pairs for each event.  
The resulting distributions of the three metrics are shown in Fig.~\ref{fig:metric_summary}. %
Based on visual inspection of representative cases, we adopt the following empirical thresholds as indicators of significant waveform-model differences:
\begin{align}
    \mathrm{JS} &\ge 0.05, \nonumber \\
    \Delta &\ge 0.3, \nonumber \\
    1 - f_{\mathrm{overlap}} &\ge 0.2. 
\end{align}
These values roughly correspond to either a clearly visible shift of the posterior mode ($\Delta\simeq 0.3$), a 20\% or larger reduction in overlap of the 90\% credible intervals, or an information-theoretic distance $\mathrm{JS}\gtrsim0.05$, for which posteriors could be distinctly separated in one or more parameters.  
Events exceeding any of these thresholds are flagged as potentially affected by waveform-model systematics. According to these criteria, we identify seven such events: GW230628\_231200, GW230814\_230901, GW230927\_153832, GW231028\_153006, GW231118\_090602, GW231123\_135430, GW231226\_101520. 
GW231028\_153006, GW231118\_090602 and GW231123\_135430 were also flagged as events with potential systematics in \cite{LIGOScientific:2025slb}, together with GW230624\_113103 and GW231118\_005626, which do not exceed any of our thresholds of metrics for systematics, since they show only tiny deviations in the posteriors shape.
We also note that GW231123\_135430 is identified in the next section, Sec.~ \ref{sec:results_eccentricity}, as an eccentric candidate. Some of the signal properties of the events that show waveform systematics are listed in table~\ref{tab:ordered}, together with other events of interest.

\subsection{Dependence of systematics on source properties}

\begin{figure}[t!]
    \centering
    \includegraphics[width=\linewidth]{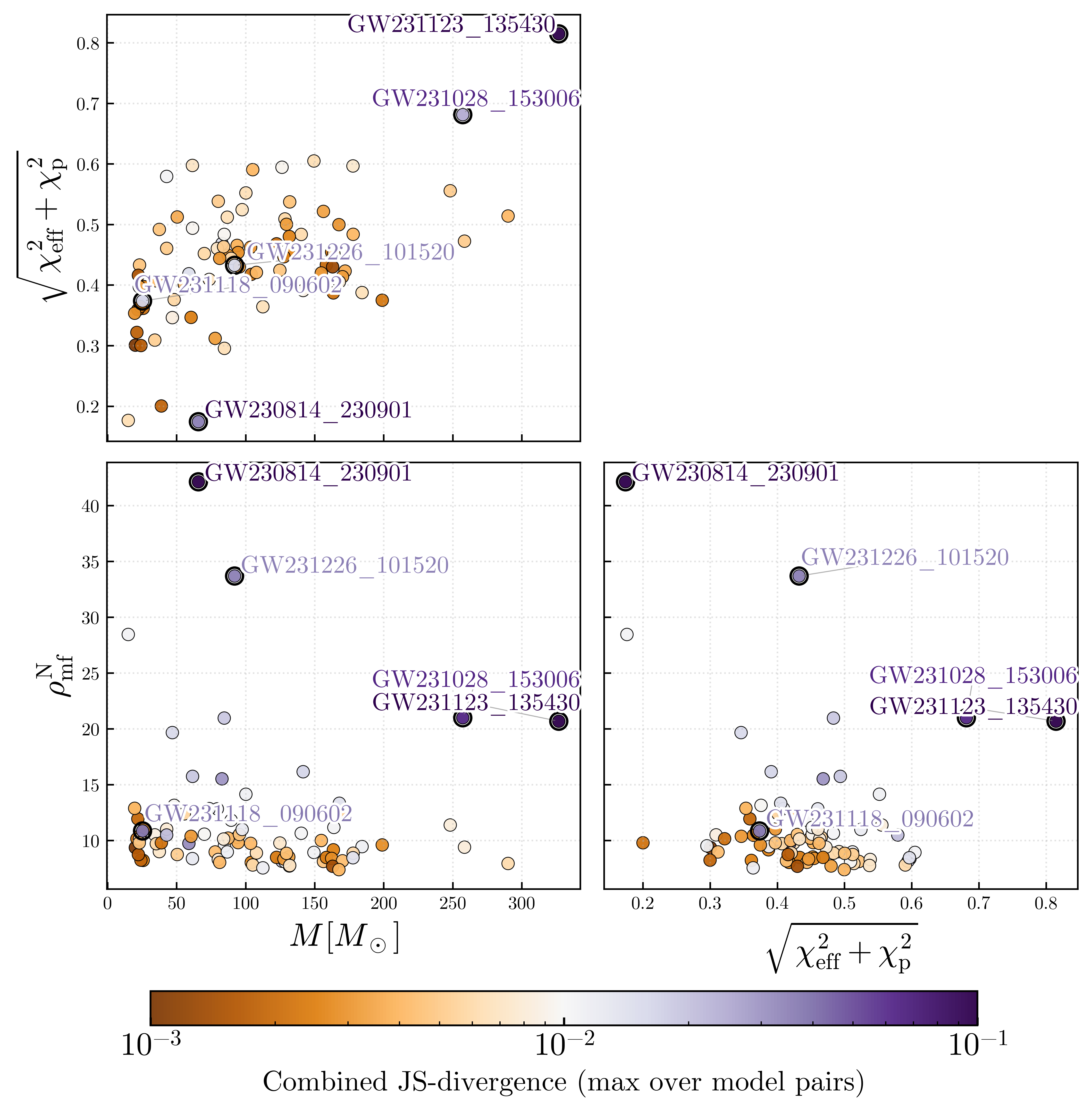}
    \caption{
        Dependence of waveform-model systematics on source parameters.  
        Each point corresponds to a \GWTCFOUR event, positioned according to the median posterior values of the redshifted total mass $M$, the combined spin magnitude $\sqrt{\chi_{\rm eff}^{2} + \chi_{\rm p}^{2}}$, and the network matched-filter signal-to-noise ratio $\rho_{\rm mf}^{N}$ obtained from the reference \texttt{GWTC} analysis (using the combined samples).  
        Points are colored by the combined JS divergence (maximized over all model pairs and parameters), shown in logarithmic scale.  
        Events with the largest JS values are annotated.  
        The figure shows that only a small subset of high-mass, high-spin, or high-SNR systems exhibit significant waveform-model differences, while most of the population occupies the region where posteriors are consistent across models.
    }
    \label{fig:triangular_JS}
\end{figure}

We further investigate the origin of the observed discrepancies by studying the dependence of the JS divergence on the source properties. 
In Fig.~\ref{fig:triangular_JS}, we show the location of each event in the three-dimensional space $(M, \sqrt{\chi_{\rm eff}^2+\chi_{\rm p}^2}, \rho_{\rm mf}^N)$ using the median values from the combined samples released in \GWTCFOUR, where $\sqrt{\chi_{\rm eff}^2+\chi_{\rm p}^2}$ is a simple combined spin parameter that accounts for spin-precessing effects (notice that since $\chi_{\rm p}$ is poorly measured for most events, this combined parameter has a shift toward higher values due to prior dominance), and $\rho_{\rm mf}^N$ denotes the network matched-filter signal-to-noise ratio. The color gradient indicates the maximum combined JS divergence between \IMRPhenomTPHM and \IMRPhenomXPNR and all the available models in \GWTCFOUR: \IMRPhenomXPHMST, \SEOBNRFIVEPHM, \NRSur and \IMRPhenomXOFOURa.
Most events cluster in a ``vanilla'' region of moderate mass, spin, and SNR, where the different waveform models yield fully consistent results.  
Larger JS divergences are observed only for systems with higher total masses, higher SNRs, or larger spins, corresponding to the regions where waveform modeling is not (well) informed by \nr{}, due to the sparsity of simulations, or where precession effects are stronger.
One notable exception, GW231118\_090602, lies in the low-mass, low-SNR region but is known to be affected by a data glitch, as noted in Tables~\ref{table:DQ} and ~\ref{tab:ordered}.

\subsection{Posterior comparison for events with potential systematics}

\begin{figure*}[t!]
    \centering
    \includegraphics[width=\linewidth]{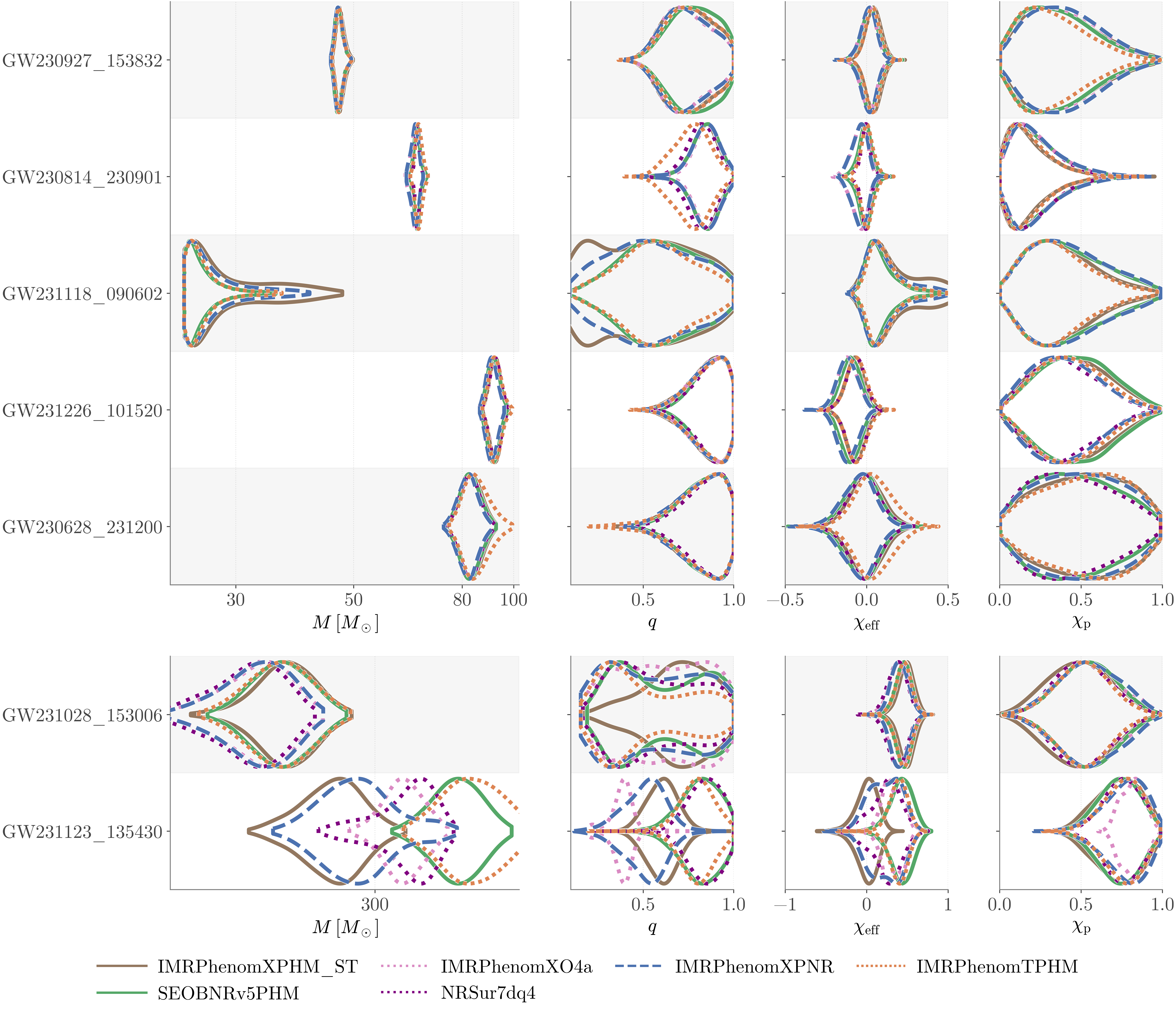}
    \caption{
        Comparison of posterior distributions for the seven events flagged as potentially affected by waveform–model systematics.  
        Each column corresponds to one intrinsic parameter ($M$, $q$, $\chi_{\rm eff}$, and $\chi_{\rm p}$), and each row to a different event.  
        Posteriors are shown for \IMRPhenomXPNR, \IMRPhenomTPHM, and the models used in GWTC–4. The majority of these events only show mild discrepancies in the inferred parameters, while the most massive ones (GW231028\_153006 and GW231123\_135430) exhibit visible larger discrepancies.
    }
    \label{fig:violin_systematics}
\end{figure*}

Figure~\ref{fig:violin_systematics} compares the posterior distributions of the seven events identified as potentially affected by waveform-model systematics according to the metrics discussed above: 
GW230628\_231200, GW230814\_230901, GW230927\_153832, GW231028\_153006, GW231118\_090602, GW231123\_135430, and GW231226\_101520.  
In most cases, the differences between models are moderate and confined to one or two intrinsic parameters, typically the total mass or effective spin, while the overall posterior structure remains consistent across models.

\begin{figure}[t!]
    \centering
    \includegraphics[width=0.99\columnwidth]{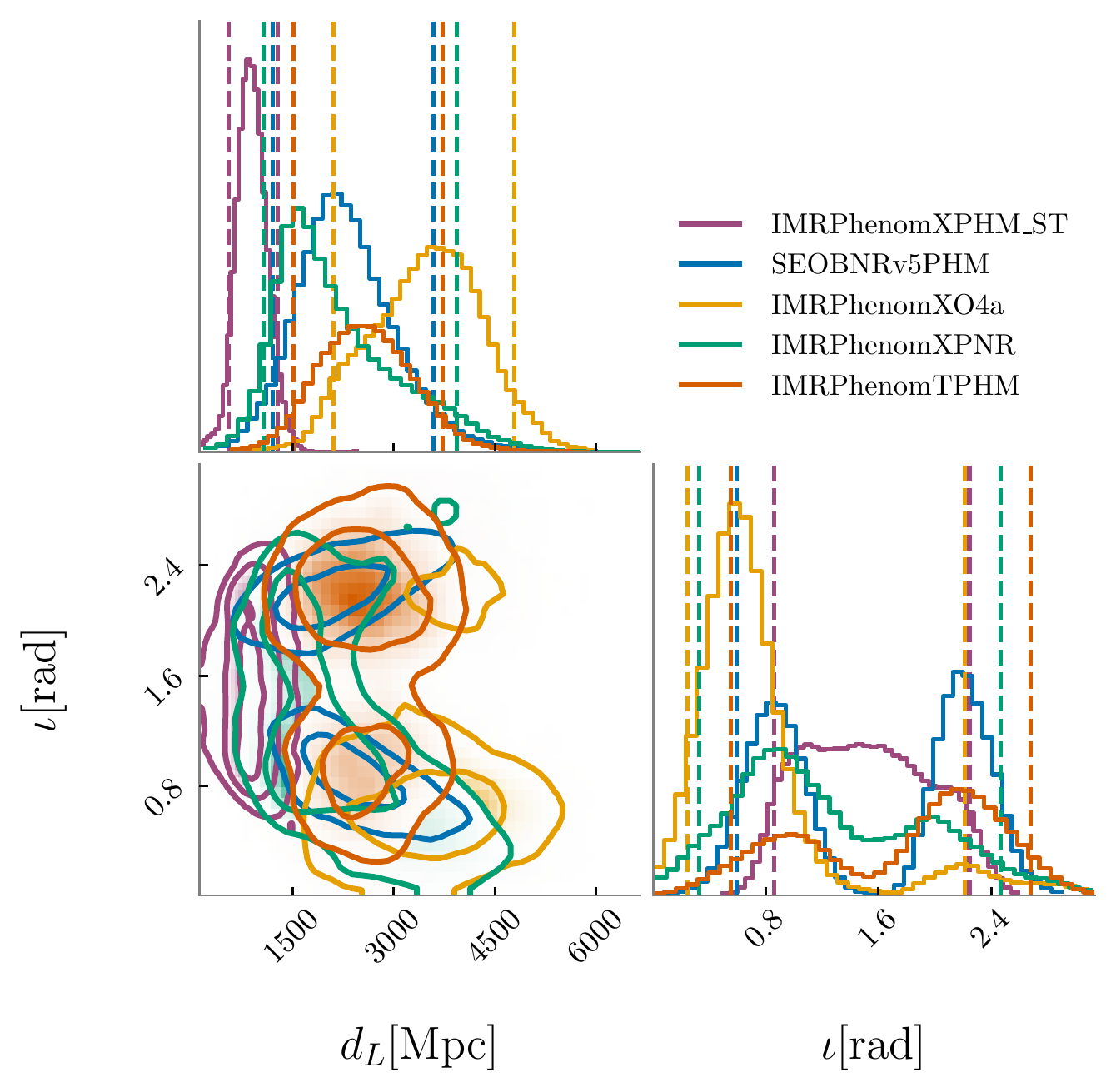}
    \caption{
        Inferred luminosity distance and source inclination for GW231123\_135430 using different waveform models. The discrepant results of  \IMRPhenomXPHMST are not reproduced by \IMRPhenomXPNR nor \IMRPhenomTPHM, which show improved consistency in these parameters with
        the other models.
    }
    \label{fig:corner_231123_dist_iota}
\end{figure}

Two events, GW231028\_153006 and GW231123\_135430, stand out as the most discrepant cases.  
Both correspond to high-mass systems ($M \gtrsim 200\,M_\odot$), with support for spinning black-holes (evidence of positive $\chi_{\rm eff}$ and a posterior distribution for $\chi_{\rm p}$ distinct from the prior towards larger values) and SNR distributions above 20 at 90\% credibility. Together, these characteristics place these events in a region where subdominant modes, ringdown modeling and spin-precession modeling play an enhanced role, leading to larger waveform-model differences.  
In these cases, the posteriors inferred with \IMRPhenomXPNR and \IMRPhenomTPHM show noticeable shifts relative to each other and to previous \GWTCFOUR results, consistent with the behavior observed in the systematic metrics of Fig.~\ref{fig:metric_summary}. In particular, GW231123\_135430 displayed significant systematic differences in the redshifted total mass, and our new runs continue to show this discrepancy, with \IMRPhenomTPHM showing better consistency with \SEOBNRFIVEPHM, and \IMRPhenomXPNR agreeing more with \IMRPhenomXPHMST. Nevertheless, \IMRPhenomXPHMST displayed the most significant difference in the recovery of the luminosity distance and inclination of the source, while the improved \IMRPhenomXPNR is more consistent with the rest of the models in these parameters, including \IMRPhenomTPHM, as we show in Fig.~\ref{fig:corner_231123_dist_iota}.

\begin{figure}[t!]
    \centering
    \includegraphics[width=0.99\columnwidth]{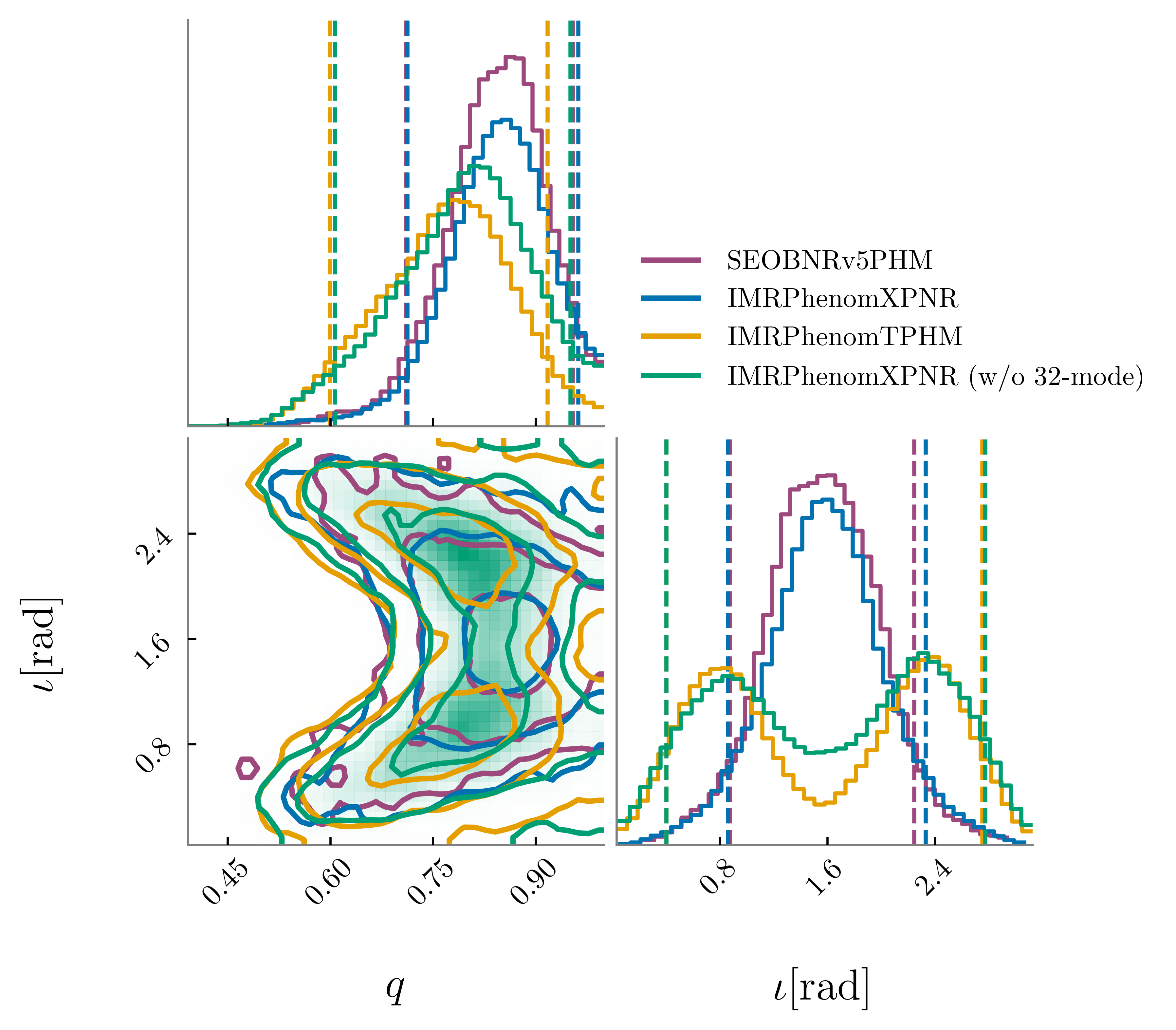}
    \caption{
        Posterior comparison for GW230814\_230901 using different waveform models.  
        The differences between \IMRPhenomXPNR and \IMRPhenomTPHM arise from the absence of the $(3,2)$ mode in \IMRPhenomTPHM.  
        When the $(3,2)$ mode is removed from \IMRPhenomXPNR, the resulting posteriors match those of \IMRPhenomTPHM, confirming the mode–content origin of the discrepancy.
    }
    \label{fig:corner_GW230814_32mode}
\end{figure}

For GW230814\_230901, the \IMRPhenomTPHM analysis predicts noticeably different total-mass, mass-ratio and inclination posteriors compared to \IMRPhenomXPNR. This event is a loud single-detector signal (with SNR above 41 with 90\% credibility) where spin effects are constrained to be small, with comparable masses, and therefore we expect better agreement between different waveform models.
We investigated this discrepancy and found that it originates from the absence of the $(3,2)$ mode in \IMRPhenomTPHM.  
When re-analyzing the event with \IMRPhenomXPNR after explicitly removing the $(3,2)$ contribution, we reproduce the \IMRPhenomTPHM results, confirming the mode-content origin of the difference. As seen in Fig.~\ref{fig:corner_GW230814_32mode}, the models (IMRPhenomXPNR and SEOBNRv5PHM) that include the $(3,2)$ harmonic, which has peaks at the poles and equator, are able to correctly measure the inclination of the source, which in this case is seen approximately edge-on, and the mass-ratio. 

We additionally find that \IMRPhenomTPHM GW230628\_231200 leads to different results than the rest of the models for GW230628\_231200, in particular the $\chi_{\rm eff}$ distribution is shifted to slightly higher values and the redshifted total mass has an extended tail towards higher masses. The remaining events ( GW230927\_153832, GW231118\_090602
\footnote{\label{footnote:GW231118}For GW231118\_090602 we 
present results with faster sampling settings due to the computational cost, in particular {\tt nlive=300} and {\tt naccept=10}, but results will be updated with our standard settings.}, and GW231226\_101520) exhibit only mild variations among waveform models, typically within the statistical uncertainties.  
In particular, GW231118\_090602 displays different levels of multimodality in some of the posterior distributions depending on the waveform model, as discussed in Sec.~\ref{sec:multimodality}.

Overall, the new \IMRPhenomXPNR and \IMRPhenomTPHM results show excellent consistency with the \GWTCFOUR analyses.  
Only a small subset of events display measurable waveform-model systematics, all of which can be traced to either extreme source properties (very high masses or spins) or data-quality effects.

\begin{figure*}[t!]
    \centering
    \includegraphics[width=\linewidth]{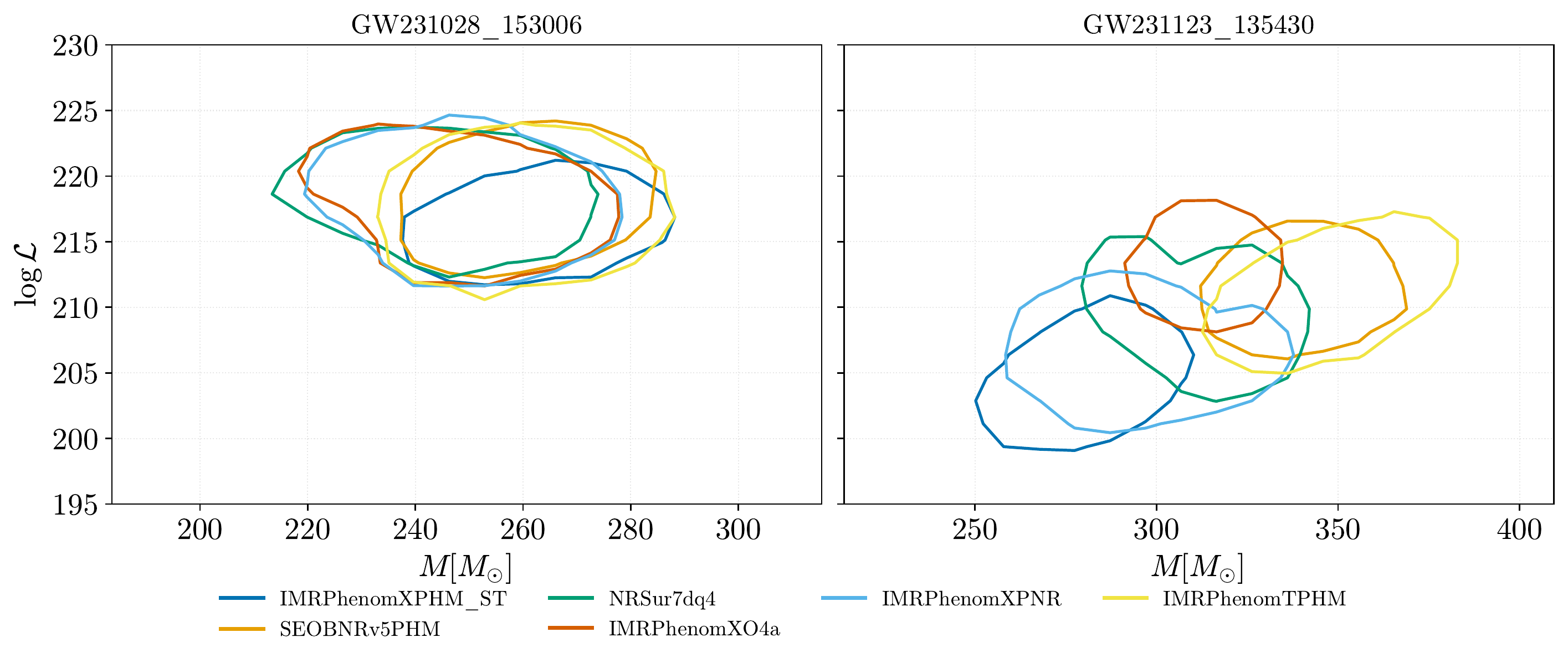}
    \caption{
        Comparison of two-dimensional posterior distributions for total mass and likelihood for the two high-mass events identified with waveform systematics in this section and in GWTC-4.0. Solid lines denote the 90\% credible level of the two-dimensional distributions.
    }
    \label{fig:log_likelihood_highmass}
\end{figure*}

We recommend the employment of combined samples including \IMRPhenomTPHM and \IMRPhenomXPNR and the available models in \GWTCFOUR for all the events not flagged in this section. For GW230814\_230901, we recommend to exclude \IMRPhenomTPHM, due to the missing mode content affecting the results. The two high-mass events display several groups of results, complicating the recommendation of specific waveform models. We inspect the likelihood recovery of the several results for these events in Fig.~\ref{fig:log_likelihood_highmass}, showing that \IMRPhenomXPHMST systematically recovers less likelihood (and therefore SNR), consistent with the modeling limitations of the merger-ringdown for spin-precessing systems in this approximant (despite these limitations, \IMRPhenomXPHMST consistently identify these systems as high-mass and spinning, showing its adequacy for efficient scanning of events). We therefore recommend not employing the results from this model for these events. For GW231028\_153006, the models with calibration to spin-precessing NR simulations (including the dominant equatorial-asymmetry effect), \NRSur, \IMRPhenomXOFOURa and \IMRPhenomXPNR, produce a different result than the uncalibrated models, which suggest the relevance of this calibration for describing this event (despite recovering similar likelihood distributions to the uncalibrated models, as seen in Fig.~\ref{fig:log_likelihood_highmass}, the employment of the calibrated models is recommended due to its improved physical completeness).
For GW231123\_135430, three main groups of results are present, with all the time-domain models and \IMRPhenomXOFOURa showing similar likelihood posteriors (see Fig.~\ref{fig:log_likelihood_highmass}), while \IMRPhenomXPHMST and \IMRPhenomXPNR are bounded to lower likelihood values, suggesting that they fit slightly worse the data. For the other events identified with systematics effects, since differences in the posterior distributions are small, a conservative recommendation would be to employ all the available models.

\subsection{Events with multimodal posteriors in \GWTCFOUR}\label{sec:multimodality}

As discussed in \cite{LIGOScientific:2025slb}, multimodality in the inferred posterior distributions can occur when the inferred parameters fall in regions where waveform models develop intricate structure, or due to noise features, in particular for low-SNR signals.
In such situations, small differences in parameters can shift support between competing likelihood maxima, resulting in multiple posterior modes.
Five events have been found to show multimodal posteriors for individual waveform models in \GWTCFOUR: GW230712\_090405, GW230723\_101834, GW231028\_153006, GW231118\_090602 and GW231123\_135430, the latter three flagged as displaying waveform systematics both in the \GWTCFOUR analysis and in our analysis. Here we discuss our updated results for these events, not finding any additional multimodal event arising from our new results with \IMRPhenomXPNR and \IMRPhenomTPHM.

GW230712\_090405 displayed multimodalities in the inferred redshifted chirp mass for the \GWTCFOUR analyses, a feature still present in our new quasi-circular results. Nevertheless, the multimodality disappears for the eccentric analysis presented in Sec.~\ref{sec:results_eccentricity} (see Fig.~\ref{fig:multimodal_posterior_gw230712}), where this event stands out as an eccentric candidate with the highest inferred eccentricity and a high Bayes factor favouring the eccentric hypothesis. Regarding the quasi-circular analysis, the \IMRPhenomXPNR results show a similar trimodal distribution as the one inferred by \IMRPhenomXOFOURa, which is correlated with a secondary mode higher-mass mode in the redshifted secondary mass. The results from \IMRPhenomTPHM enhance the middle-mass mode while suppresses the higher-mass one, resulting in a bimodal distribution for the redshifted chirp mass, more consistent with the results from \SEOBNRFIVEPHM.

GW230723\_101834 displayed also a trimodal redshifted chirp mass distribution for the \IMRPhenomXPHMST results in \GWTCFOUR, with two small modes at lower and higher values of the prominent central mode. Our new results both for \IMRPhenomXPNR and \IMRPhenomTPHM suppress the high-mass mode, leaving the secondary low-mass mode also present in the \SEOBNRFIVEPHM analysis from \GWTCFOUR. Additionally, a secondary mode in $\chi_{\rm eff}$ present in \IMRPhenomXPHMST is suppressed also for our \IMRPhenomXPNR and \IMRPhenomTPHM, in agreement with \SEOBNRFIVEPHM, since this mode was correlated with the high-mass mode in the chirp mass.

GW231028\_153006 displayed trimodal mass-ratio and secondary mass distributions, bimodal primary mass distribution and a secondary lower chirp mass mode for \IMRPhenomXOFOURa, and bimodal mass-ratio and individual mass distributions for \SEOBNRFIVEPHM in \GWTCFOUR. Our results with \IMRPhenomTPHM show consistency with \SEOBNRFIVEPHM, with a bimodal mass-ratio distribution enhancing the more asymmetric mode. While our \IMRPhenomXPNR results also enhance this mode (and suppresses the middle mass-ratio mode of \IMRPhenomXOFOURa), they display an additional mass-ratio mode at even lower mass-ratio values (more asymmetric masses), correlated with high primary-mass and low secondary-mass modes.

GW231118\_090602 results with \IMRPhenomXPHMST in \GWTCFOUR showed a bimodal secondary mass distribution correlated with a small secondary mode in the primary mass, which together enhanced a low mass-ratio mode not present in the \SEOBNRFIVEPHM analyses. Our \IMRPhenomXPNR results suppress this secondary mode, but still show a wider mass distribution going to lower secondary mass and higher primary mass, therefore constraining less the mass-ratio of the system. Our \IMRPhenomTPHM results do not show this feature, contraining the mass parameters at the same level than \SEOBNRFIVEPHM.

GW231123\_135430 results with \IMRPhenomXOFOURa in \GWTCFOUR displayed an additional mode in the secondary mass (resulting in additional low mass-ratio and low chirp-mass modes, as well as a bimodal $\chi_{\rm eff}$ distribution), a feature that is absent in our \IMRPhenomTPHM results but is reversed in our \IMRPhenomXPNR results, which show an additional higher secondary-mass mode.

In addition to multimodal posteriors for individual models, GW231028\_153006 and GW231123\_135430 displayed additional multimodalities in the combined samples presented in \GWTCFOUR. As we have seen in the previous section, these two high-mass event show several groups of results for the different models, therefore producing a multimodal combined distribution. Our results do not resolve this issue, since \IMRPhenomXPNR and \IMRPhenomTPHM results prefer a different group for both events.

\section{Eccentric Analysis}
\label{sec:results_eccentricity}

In this section we analyze the GW events released in the \GWTCFOUR catalog~\cite{LIGOScientific:2025slb} 
with  the eccentric multipolar aligned-spin phenomenological \phTE model~\cite{Planas:2025feq}. 
We compare the eccentric results with the ones obtained using the multipolar QC aligned-spin \phTHM model~\cite{Estelles:2020osj,Estelles:2020twz}, and the multipolar QC precessing-spin \phTPHM model~\cite{Estelles:2021gvs}, whose results are presented in more detail in Sec.~\ref{sec:results:QC}, and compute Bayes factors between the different hypotheses.

\subsection{Methodology}

The analysis performed in this section relies on multipolar time-domain waveforms. In order to ensure that all the waveform modes are in band at the starting frequency of likelihood integration $f^{\mathcal{L}}_{\rm{start}}$, the generation of time-domain waveforms has to start at a time corresponding to an earlier frequency $f^{\rm{wf}}_{\rm{min}}$. This frequency scales as the inverse of $m_{\rm{max}}$, the maximum $m$-index in the list of multipoles included in the waveform (see \cite{Ursell:2025ufb} for biases incurred otherwise):
\begin{equation}
f^{\rm{wf}}_{\rm{min}}=2f^{\mathcal{L}}_{\rm{start}}/m_{\rm{max}}.
\label{eq:eqFreq}
\end{equation}
The $(5,\pm5)$ harmonic of \IMRPhenomTEHM would then require to lower the start frequency of the waveform 
from 20~\si{Hz} to 8~\si{Hz}. Given the small contribution of this mode we instead start waveform
generation at 10~\si{Hz} in order to reduce the computational cost of the analysis. In some low-mass events mentioned in Sec.~\ref{sec:results_eccentricity:results}, the $(5,\pm5)$-mode is not included to further reduce the computational cost.

Before presenting the results of our analysis with \IMRPhenomTEHM, it is worth briefly discussing the definition of the eccentricity parameter, which is not uniquely defined in general relativity (GR). As a consequence, waveform models may define eccentricity in terms of orbital elements, the trajectories of the compact objects, or quantities such as the system’s energy and angular momentum. In particular in Ref.~\cite{Ramos-Buades:2022lgf} it was proposed to the use an eccentricity measure extracted from the gravitational waveform itself, which was later implemented in a highly optimized python package \texttt{gw-eccentricity}~\cite{Shaikh:2023ypz}. This GW eccentricity, $e_{\rm gw}$, is computed from the maxima and minima of the (2,2)-mode amplitude or frequency, and similarly a GW mean anomaly, $l_{\rm gw}$, can be computed~\cite{Ramos-Buades:2022lgf,Shaikh:2023ypz}. We will comment on the comparison between our results and the ones obtained with \texttt{gw-eccentricity} in Sec.~\ref{sec:results_eccentricity:discussion}.

\begin{figure*}[t!]
    \centering
    \includegraphics[height=\textheight]{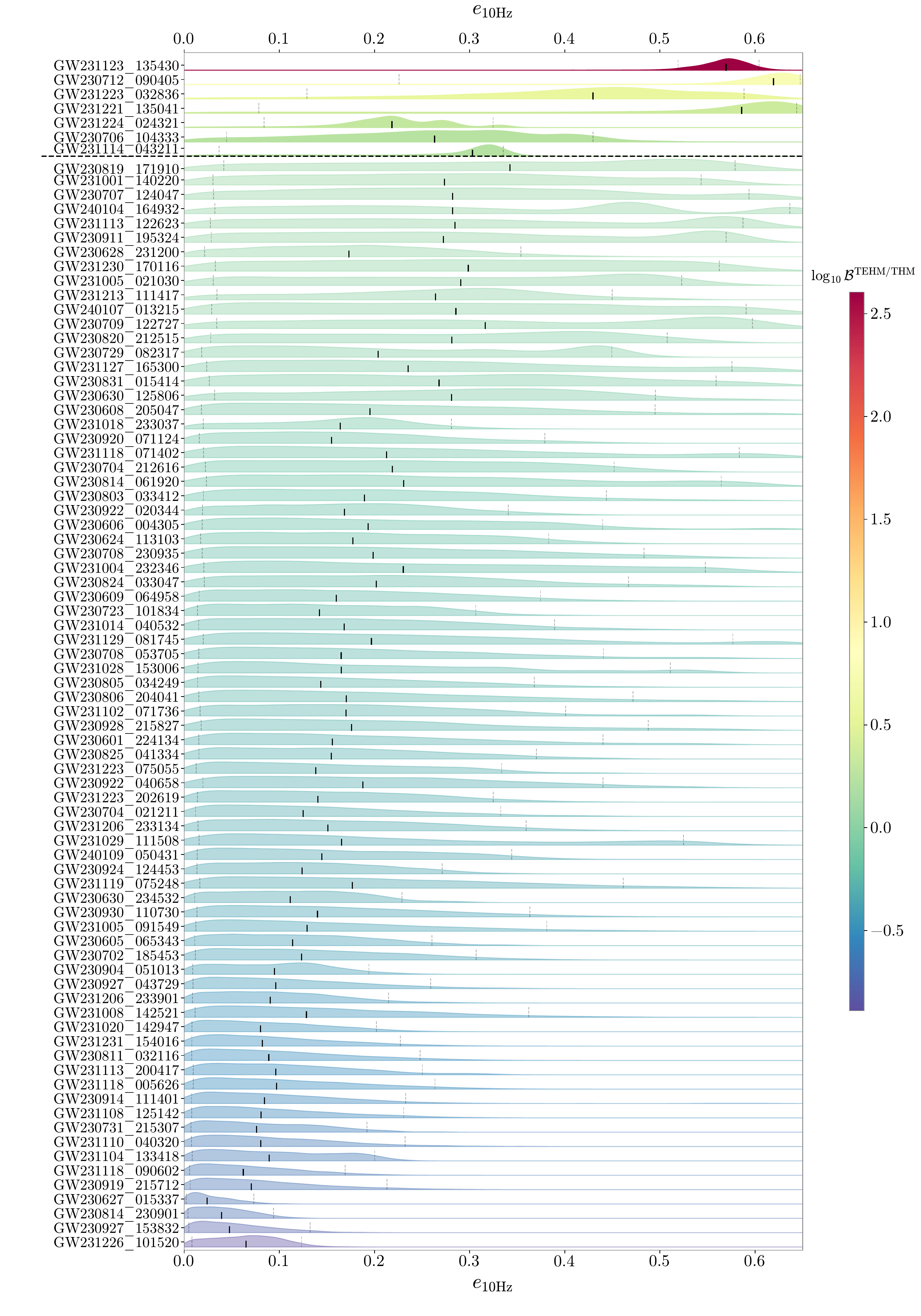}
    \caption{
     Marginal probability distribution for the orbital eccentricity at a reference frequency of 10 Hz produced for the O4a GW events considered in this work. The eccentricity has been measured using the eccentric multipolar phenomenological \phTE model~\cite{Planas:2025feq}. In the y-axis the GW events have been ordered by increasing value of the log-10 Bayes factor against the quasicircular aligned-spin model \phTHM. Additionally, the posteriors are color coded by their value of log-10 Bayes factor. The thick black lines in the posteriors correspond to their median values, while the vertical dashed lines indicate the $90\%$ credible intervals of each distribution.
     Note that due to the different sampling settings employed, we do not include GW230627\_015337 here (see footnote~\ref{footnote:GW230627} for more details).
    }
    \label{fig:all_ecc_posteriors}
\end{figure*}

\begin{figure*}[t!]
    \centering
    \includegraphics[width=\textwidth]{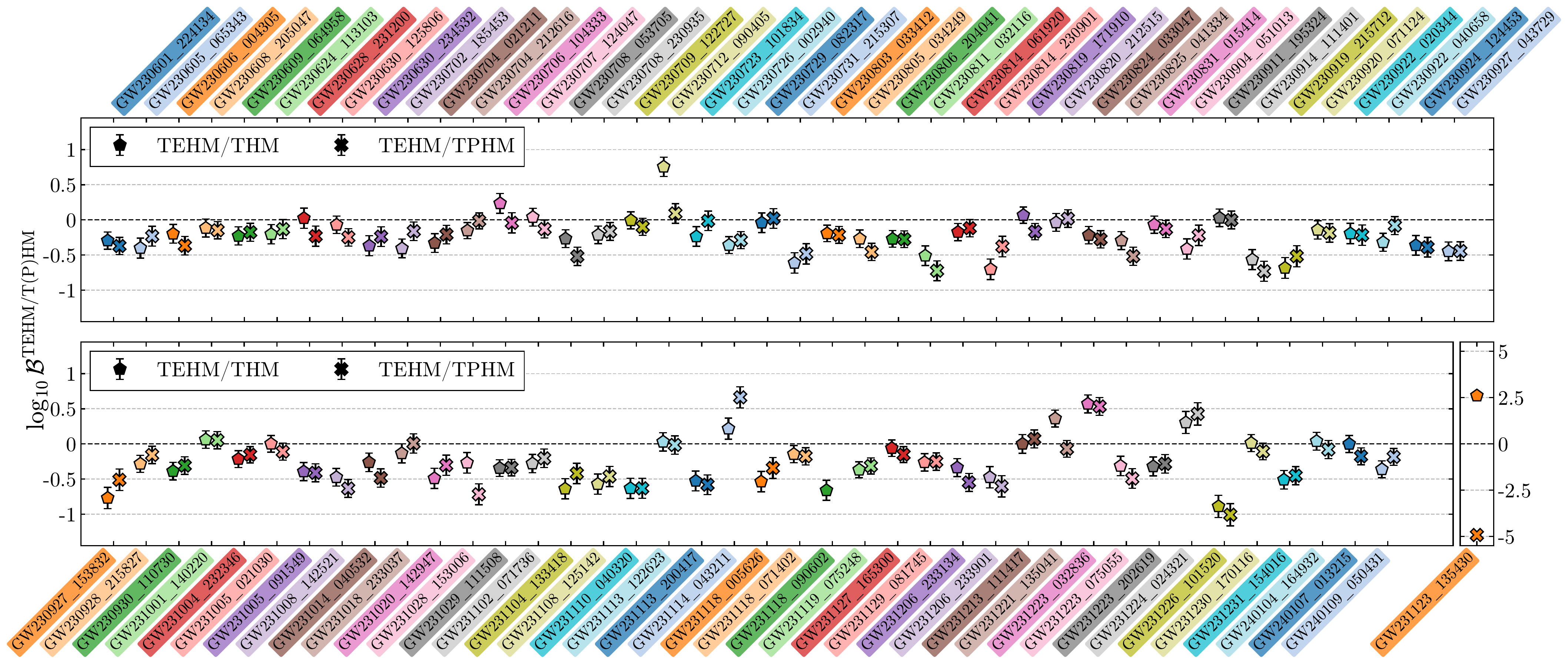}
    \caption{
    Log-10 Bayes factors comparing the eccentric aligned-spin hypothesis computed with the \phTE model \cite{Planas:2025feq} against the aligned quasi-circular hypothesis computed with the \phTHM model \cite{Estelles:2020osj,Estelles:2020twz} (TEHM/THM), as well as the comparison of the eccentric aligned-spin hypothesis against the QC precessing-spin one computed with the \phTPHM model \cite{Estelles:2021gvs} (TEHM/TPHM)  for each O4a GW events considered in this work. Each nominal value of log-10 Bayes factor contains its associated error. In order to ease the visualization of the results the events have been split into two panels, each one with a horizontal dashed line marking the zero value, while \text{GW231123\_125430} is displayed in a separate inset.  
    Note that due to the different sampling settings employed, we do not include GW230627\_015337 and the \IMRPhenomTPHM results for GW231118\_090602 in this comparison (see footnotes~\ref{footnote:GW231118} and~\ref{footnote:GW230627} for more details).
    }
    \label{fig:bayes_factor_e_qc}
\end{figure*}

\subsection{Results}\label{sec:results_eccentricity:results}

Figure~\ref{fig:all_ecc_posteriors} shows the one-dimensional marginal posterior probability distribution of the orbital eccentricity measured at 10~\si{Hz} for 83 GW events.\footnote{\label{footnote:GW230627}Due to computational cost for GW230627\_015337 we perform parameter estimation with lower sampling settings with \IMRPhenomTEHM, which do not show hints of eccentricity. For this specific event, parameter estimation runs with a faster IMR frequency-domain eccentric model in development~\cite{XEpaper} also do not show signs of eccentricity. We decided to not include this event in this section since log-10 Bayes factors could yield unfaithful values for different sampling settings.}
In Fig.~\ref{fig:all_ecc_posteriors} the events are listed from top to bottom in decreasing order of log-10 Bayes factor between the eccentric aligned-spin hypothesis calculated with the \phTE model~\cite{Planas:2025feq}, and the QC aligned-spin hypothesis computed with the \phTHM model \cite{Estelles:2020osj,Estelles:2020twz}, $\log_{10}\mathcal{B}^{\rm{TEHM/THM}}$, which is indicated by the color of each probability distribution.
The thick black vertical lines represent the median and the dashed gray vertical lines mark the 90\% credible interval.
The events above the dashed black horizontal line have values of $\log_{10}\mathcal{B}^{\rm{TEHM/THM}} - \delta \rm{error}>0$, where $\delta \rm{error}>0$ is the statistical error in the calculation of the Bayes factor obtained from the error in the evidences computed by the \texttt{dynesty} nested sampler~\cite{2020MNRAS.493.3132S,sergey_koposov_2025_17268284}.
For this work, we adopt this criterion as the metric to identify interesting events, which could hint of signatures of eccentricity.
Alternative metrics, such as the median of the one-dimensional eccentricity posterior normalized by its variance, were also tested and yielded largely consistent candidates.

Most events, as shown in Fig.~\ref{fig:all_ecc_posteriors}, have negative log-10 Bayes factors and eccentric posteriors that include substantial support for  $e_{\rm{10\si{Hz}}}=0$ or span the entire prior range,
thus being consistent with QC mergers. 

In the following, we describe the main properties of events for which the eccentric aligned-spin (\IMRPhenomTEHM) hypothesis is statistically consistent with positive values of log-10 Bayes factor against the aligned-spin QC (\IMRPhenomTHM) hypothesis. For each candidate, we compare the network matched-filter SNR obtained with both waveform models (which we define as $\rm{SNR}^{TEHM/THM}$ for simplicity).\footnote{Note that the network matched-filter SNR values reported in this section need to be postprocessed in order to be compared to the ones reported in \GWTCFOUR, since the \BILBY versions and sampling rates do not match (see App.~\ref{app:likelhood} for more details).} For the rest of the section, in order to facilitate comparisons, all quoted parameters have been extracted from the \GWTCFOUR~\cite{LIGOScientific:2025slb} catalog unless stated otherwise.

\subsubsection{GW231123\_135430}

GW231123\_135430 is the most massive event detected until O4a with source frame component masses $137^{+22}_{-17}M_\odot$ and $103^{+20}_{-52}M_\odot$, and high spins $0.90^{+0.10}_{-0.19}$ and $0.80^{+0.20}_{-0.51}$ according to the LVK analysis~\cite{LIGOScientific:2025rsn}. Two additional relevant features of the analysis of this event are the use of glitch-subtracted data in order to mitigate the impact of some noise transients in LIGO-Hanford and LIGO-Livingston at the time of the event\footnote{The glitch observed in LIGO-Livingston had no impact on the GW231123\_135430 analysis. Therefore glitch-subtraction techniques were exclusively applied to LIGO-Hanford data, while the LIGO-Livingston data was unmodified. (See Sec.~III of~\cite{LIGOScientific:2025rsn} for further details on data quality for this event).}, and the presence of waveform systematics among QC precessing-spin waveform models (see Sec.~\ref{sec:results:QC} for a detailed discussion).

Based on our study, GW231123\_135430 shows the highest preference for the aligned-spin eccentric hypothesis against the QC aligned-spin hypothesis with a $\log_{10}\mathcal{B}^{\rm{TEHM/THM}}=2.60^{+0.15}_{-0.15}$, and a corresponding high value of the orbital eccentricity $e_{10\si{Hz}}=0.57^{+0.03}_{-0.05}$ (median and 90\% credible intervals).
Regarding the network matched filter SNR, it is slightly larger for \IMRPhenomTEHM, $\rm{SNR}^{TEHM}=19.9^{+0.2}_{-0.3}$, than for \IMRPhenomTHM, $\rm{SNR}^{THM}=19.4^{+0.2}_{-0.3}$.
However, such evidence for eccentricity vanishes when comparing against the QC precessing spin hypothesis with $\log_{10}\mathcal{B}^{\rm{TEHM/TPHM}}=-4.92^{+0.16}_{-0.16}$, as shown in Fig.~\ref{fig:bayes_factor_e_qc}. 
These results can be explained by the importance of spin-precession in this event, which is not included in the \phTE model used for the analysis,  and by the short duration of the event: only $\sim 5$ cycles are observed in band, which challenges the analysis, as most of the state-of-the-art eccentric IMR models~\cite{Gamboa:2024hli, Gamba:2024cvy, Planas:2025feq} (including \phTE) assume circularization of the binary at merger-ringdown, which substantially complicates the measurement of eccentricity for high redshifted mass events.

\begin{figure}[t!]
    \centering
    \includegraphics[width=\columnwidth]{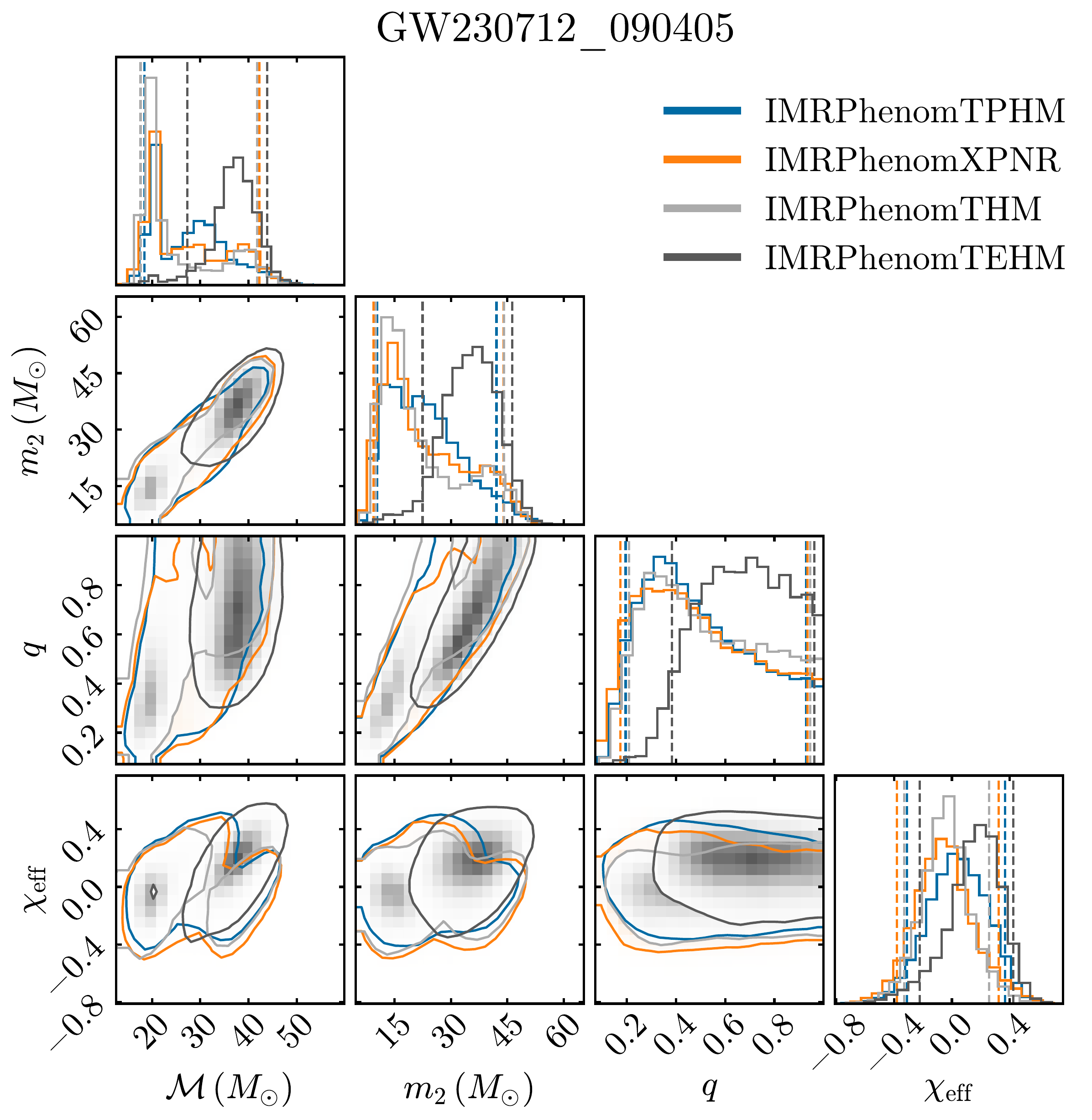}
    \caption{
     Comparison between some marginalized one- and two-dimensional posterior probability distributions obtained for GW230712\_090405 with the QC models \IMRPhenomTPHM (blue), \IMRPhenomXPNR (orange), and \IMRPhenomTHM (light gray), and the eccentric model \IMRPhenomTEHM (dark gray). 
     We show the posterior distributions for the redshifted chirp mass, $\mathcal{M}$, redshifted secondary mass, $m_2$, mass ratio, $q$, and the effective inspiral spin, $\chi_{\rm{eff}}$.
     The dashed vertical lines in the one-dimensional posteriors represent the $90\%$ credible intervals of each distribution.
    }
    \label{fig:multimodal_posterior_gw230712}
\end{figure}
    
\subsubsection{GW230712\_090405}

GW230712\_090405 was found, according to the LVK analysis~\cite{LIGOScientific:2025slb}, with the coherent Wave Burst (cWB) search pipeline with a higher coherent SNR than the one obtained with matched-filtering techniques, which, among other effects, could indicate hints of non-captured physics~\cite{PhysRevD.111.023054}, such as orbital eccentricity, in the matched-filter templates used by the LVK~\cite{LIGOScientific:2025slb}.
    The source of this candidate has a source-frame total mass $M_{\rm{src}}=46^{+22}_{-11}\,M_{\odot}$, and shows one of the greatest deviations between the posterior distribution for $\chi_{\rm{p}}$ and its effective prior (after conditioning on the $\chi_{\rm{eff}}$ posterior distributions) with an estimated $\chi_{\rm{p}}=0.61^{+0.31}_{-0.43}$. It provides support for unequal masses, discarding $q\leq0.85$ at the 90\% credible interval, and has shown to have a bimodal chirp mass posterior distribution.
    
    Our analysis with \IMRPhenomTEHM, shows that GW230712\_090405 has the largest recovered orbital eccentricity of $e_{10\si{Hz}}=0.62^{+0.03}_{-0.39}$ and that the matched-filter SNR is slightly larger, $\rm{SNR}^{TEHM}=8.6^{+0.4}_{-0.9}$, than for the rest of the QC models considered in this study (e.g. $\rm{SNR}^{THM}=8.0^{+0.6}_{-0.8}$, $\rm{SNR}^{TPHM}=8.3^{+0.6}_{-0.9}$ )
    As shown in Fig.~\ref{fig:multimodal_posterior_gw230712}, the redshifted chirp mass posterior distribution inferred by \IMRPhenomTEHM is however unimodal and peaked at the secondary chirp-mass mode recovered by the quasiciruclar IMRPhenom models see discussion for QC models in Sec.~\ref{sec:multimodality}). 
    The same behavior is observed for the redshifted secondary mass, and the recovered mass ratio is shifted toward more comparable mass systems. 
    The aligned-spin eccentric hypothesis is favored against the QC aligned-spin hypothesis with a log-10 Bayes factor of $\log_{10}\mathcal{B}^{\rm{TEHM/THM}}=0.76^{+0.14}_{-0.14}$.
    When comparing the aligned-spin eccentric hypothesis against with the QC misaligned hypothesis the log-10 Bayes factor is significantly reduced to $\log_{10}\mathcal{B}^{\rm{TEHM/TPHM}}=0.09^{+0.14}_{-0.14}$, which, as for GW231123\_135430, could be explained by the importance of spin-precession in this event.

\subsubsection{ GW231223\_032836}

GW231223\_032836 is a system with a source-frame total mass of $M_{\rm{src}}=76^{+18}_{-13}\,M_{\odot}$. At the time of this event, the LIGO-Hanford detector was affected by a noise transient, and glitch-subtraction techniques were applied to the data.
    As shown in Fig.~\ref{fig:all_ecc_posteriors}, our criterion identifies this event as an eccentric candidate with $\log_{10}\mathcal{B}^{\rm{TEHM/THM}}=0.57^{+0.13}_{-0.13}$ and  $e_{10\si{Hz}}=0.43^{+0.16}_{-0.30}$.
    It is important to note that the frequency at which the orbital eccentricity is measured (10~\si{Hz}) lies at the edge of the range ($10-25~\si{Hz}$) where the glitch-subtraction procedure was applied~\cite{LIGOScientific:2025slb}.
    The SNRs for the eccentric and QC models are $\rm{SNR}^{TEHM}=9.1^{+0.4}_{-0.6}$ and $\rm{SNR}^{THM}=8.8^{+0.3}_{-0.5}$, respectively.
    Importantly, the evidence of eccentricity is still preferred when comparing against the QC precessing hypothesis with a log-10 Bayes factor of $\log_{10}\mathcal{B}^{\rm{TEHM/TPHM}}=0.53^{+0.13}_{-0.13}$.

\subsubsection{GW231221\_135041} 

GW231221\_135041 is a binary coalescence with source-frame total mass $M_{\rm{src}}=76^{+21}_{-15}\,M_\odot$. For this event, glitch-subtraction techniques were applied to LIGO-Hanford data between 200~\si{Hz} and 450~\si{Hz}. Like GW230712\_090405, the coherent SNR obtained by cWB-BBH for this event is larger than the matched-filter SNR obtained by template-based LVK searches, which could hint at residual orbital eccentricity~\cite{PhysRevD.111.023054}.\footnote{Note that other events such as GW231004\_232346 and GW231230\_170116 also share this feature but were not identified as eccentric candidates by our study. 
This might indicate the presence of other physical effects, apart form eccentricity, that are not currently implemented in the template-based searches used by LVK, such as misaligned spins or the effect of subdominant modes. In fact, GW231004\_232346 and GW231230\_170116 are close to edge-on systems, with $\iota=1.50^{+1.23}_{-1.10}\,\si{rad}$ and $\iota=1.66^{+1.13}_{-1.30}\,\si{rad}$, respectively, which emphasize the impact of higher-order modes.}
    
    The analysis of GW231221\_135041 with the \IMRPhenomTEHM model infers an orbital eccentricity at 10~\si{Hz} of $e_{10\si{Hz}}=0.59^{+0.05}_{-0.52}$, and a preference for the aligned-spin eccentric hypothesis against the aligned-spin QC hypothesis of $\log_{10}\mathcal{B}^{\rm{TEHM/THM}}=0.36^{+0.12}_{-0.12}$. The SNRs for these two cases are $\rm{SNR}^{TEHM}=7.9^{+0.5}_{-0.7}$ and $\rm{SNR}^{THM}=7.6^{+0.3}_{-0.6}$. This evidence is however dropped when comparing against the misaligned QC hypothesis to $\log_{10}\mathcal{B}^{\rm{TEHM/TPHM}}=-0.07^{+0.12}_{-0.12}$, which could be explained by the non-negligible impact of the precessing spin component, estimated as  $\chi_{\rm{p}}=0.57^{+0.33}_{-0.41}$ in \GWTCFOUR~\cite{LIGOScientific:2025slb}.

\subsubsection{GW231224\_024321} 

GW231224\_024321 has, compared to the previous eccentric candidates, a lower source-frame total mass $M_{\rm{src}}=16.7^{+1.3}_{-0.8}\,M_{\odot}$ and redshifted total mass $M=19.7^{+1.0}_{-0.3}\,M_{\odot}$, which results in a comparatively longer inspiral in the detector band before the merger-ringdown. Another additional relevant feature is the tight constraint of the effective inspiral spin around zero, $\chi_{\rm{eff}}=-0.007^{+0.076}_{-0.058}$. For this event, no glitch-subtraction postprocessing was applied to the data. However, the low-frequency cut-off for LIGO-Hanford data was raised from the standard 20~\si{Hz} to 40~\si{Hz} to exclude contaminated data.
While the number of observed orbital cycles in the LIGO-Livingston band is $\sim 95$, the rise in the lower-frequency cut-off reduced this number to $\sim 25$ for LIGO-Hanford.
In our eccentric study, the lower limit of the chirp-mass prior was set to $6.5\,M_{\odot}$, and multipolar modes were generated in \IMRPhenomTEHM up to $\ell\leq4$, excluding the $(\ell=5, m=\pm5)$, which is expected to not affect significantly the inference process~\cite{Planas:2025jny,Planas:2025plq}.\\
Our analysis with \IMRPhenomTEHM infers a residual orbital eccentricity of $e_{10\si{Hz}}=0.20^{+0.07}_{-0.14}$ at 10~\si{Hz} and $\rm{SNR}^{TEHM}=13.1^{+0.3}_{-0.4}$ (while $\rm{SNR}^{THM}=12.7^{+0.2}_{-0.3}$).
The eccentricity posterior distribution shows multimodalities (see Fig.~\ref{fig:all_ecc_posteriors}) that have been checked to be consistent with the results given by \IMRPhenomXE~\cite{XEpaper}, an IMR frequency-domain eccentric model under development for the dominant multipole ($\ell=2,\,m=\pm2$), which is not only computationally much more efficient, but, being formulated in the frequency domain, allows an analysis where subtleties associated with time-domain waveforms are avoided, in particular conditioning the time-domain data for the Fourier transform. 
These multimodalities could originate from degeneracies with other parameters at low frequencies, which has also been observed in previous studies~\cite{Planas:2025plq, Jan:2025fps}. Although the multiple peaks in the eccentricity distribution do not trigger multimodalities in other parameters' posteriors, we find that including orbital eccentricity leads to a preference for a smaller redshifted chirp mass compared to the QC results.
As shown in Fig.~\ref{fig:bayes_factor_e_qc}, the aligned-spin eccentric hypothesis is preferred against both the aligned- and misaligned-spin QC hypotheses with log-10 Bayes factors of 
$\log_{10}\mathcal{B}^{\rm{TEHM/THM}}=0.30^{+0.16}_{-0.16}$ and $\log_{10}\mathcal{B}^{\rm{TEHM/TPHM}}=0.43^{+0.16}_{-0.16}$, respectively.
    
 \subsubsection{GW230706\_104333} 
 
 GW230706\_104333 is a BBH candidate with an inferred source-frame total mass $M_{\rm{src}}=27.9^{+4.4}_{-3.0}\,M_{\odot}$ and mass ratio $q=0.71^{+0.25}_{-0.28}$.
    No data-quality issues were reported at the time of the event in either of the LIGO detectors. \\
    Our analysis labels this event as an eccentric candidate with an estimated orbital eccentricity $e_{10\si{Hz}}=0.26^{+0.17}_{-0.22}$. The aligned-spin eccentric hypothesis (with $\rm{SNR}^{TEHM}=9.1^{+0.4}_{-0.5}$) is favored against the aligned-spin QC hypothesis (with $\rm{SNR}^{THM}=8.9^{+0.2}_{-0.5}$) with a log-10 Bayes factor of $\log_{10}\mathcal{B}^{\rm{TEHM/THM}}=0.23^{+0.14}_{-0.14}$. This evidence is strongly reduced when comparing against the precessing QC hypothesis, yielding a log-10 Bayes factor of $\log_{10}\mathcal{B}^{\rm{TEHM/TPHM}}=-0.04^{+0.14}_{-0.14}$.

\begin{figure}[t!]
    \centering
    \includegraphics[width=\columnwidth]{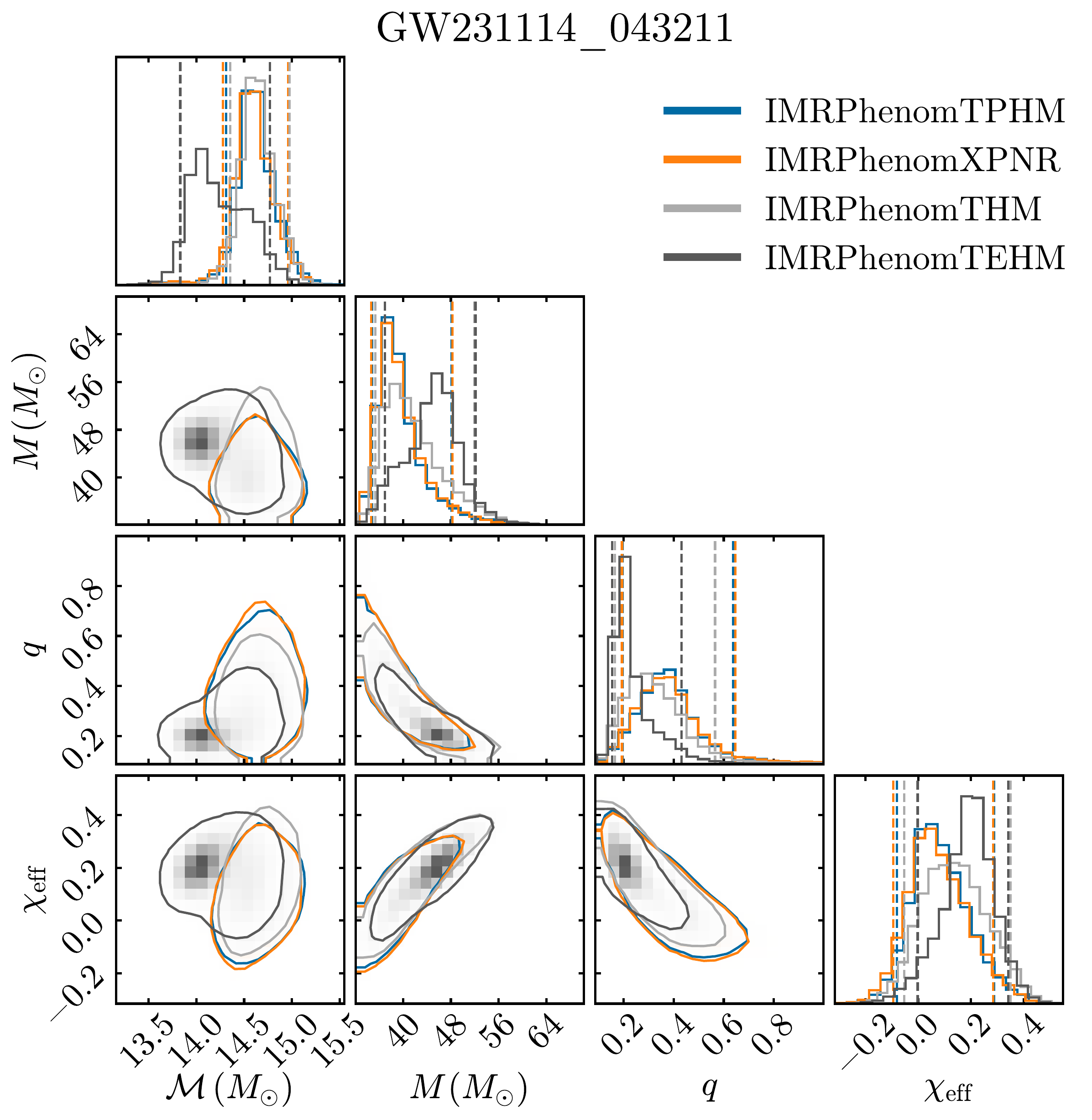}
    \caption{
     Comparison between some marginalized one- and two-dimensional posterior probability distributions obtained for GW231114\_043211 with the QC models \IMRPhenomTPHM (blue), \IMRPhenomXPNR (orange), and \IMRPhenomTHM (light gray), and the eccentric model \IMRPhenomTEHM (dark gray). 
     We show the posterior distributions for the redshifted chirp mass, $\mathcal{M}$, redshifted total mass, $M$, mass ratio, $q$, and the effective inspiral spin, $\chi_{\rm{eff}}$.
     The dashed vertical lines in the one-dimensional posteriors represent the $90\%$ credible intervals of each distribution.
    }
    \label{fig:posteriors_gw231114}
\end{figure}

\subsubsection{GW231114\_043211} GW231114\_043211 has a source-frame total mass $M_{\rm{src}}=31.0^{+7.6}_{-4.8}\,M_{\odot}$ and an effective inspiral spin of $\chi_{\rm{eff}}=0.08^{+0.22}_{-0.16}$, and shows the largest posterior support for unequal masses in \GWTCFOUR, with an inferred mass ratio of $q=0.36^{+0.27}_{-0.17}$.
    At the time of the event, LIGO-Hanford experienced a noise artifact, and glitch subtraction was performed at a frequency window from 10~\si{Hz} to 30~\si{Hz}.\\
    The result from \IMRPhenomTEHM estimates a residual orbital eccentricity of $e_{10\si{Hz}}=0.30^{+0.04}_{-0.26}$. Note that this eccentricity is measured
    at 10 Hz, which lies at the edge of the range
    where the glitch-subtraction procedure was applied. The inclusion of eccentricity, apart from increasing the SNR from $\rm{SNR}^{THM}=9.8^{+0.3}_{-0.5}$ to $\rm{SNR}^{TEHM}=10.3^{+0.6}_{-0.8}$, shows degeneracies with the redshifted chirp mass, redshifted total mass, mass ratio, and effective inspiral spin, as displayed in Fig.~\ref{fig:posteriors_gw231114}. 
   As opposed to the quasicirular models, \IMRPhenomTEHM discards $\chi_{\rm{eff}}<0$ at the 90\% credible interval ($\chi_{\rm{eff}}=0.19^{+0.15}_{-0.19}$), and the recovered mass-ratio posterior has a stronger preference for more asymmetric binaries, with an inferred value of $q=0.21^{+0.22}_{-0.06}$.
    The aligned-spin eccentric hypothesis is favored against both the aligned- and misaligned-spin quasi-circular hypotheses with $\log_{10}\mathcal{B}^{\rm{TEHM/THM}}=0.22^{+0.15}_{-0.15}$ and $\log_{10}\mathcal{B}^{\rm{TEHM/TPHM}}=0.66^{+0.15}_{-0.15}$, respectively.

\subsection{Discussion of eccentric results}\label{sec:results_eccentricity:discussion}

The eccentric candidates listed above and shown in Fig.~\ref{fig:all_ecc_posteriors} can be distributed into two groups:
events with high mass and high eccentricity $e_{10\si{Hz}}>0.4$, and events with eccentricity $e_{10\si{Hz}}<0.3$.

\subsubsection{Eccentric candidates with high eccentricity }\label{sec:results_high_eccentricity:discussion}

The first group consists of GW231123\_13543, GW230712\_090405, GW231223\_032836, and GW231221\_135041, characterized by eccentricity median values of $e_{10\si{Hz}}>0.4$.
All these events show inferred redshifted total masses with \IMRPhenomTEHM greater than $90\,M_\odot$.
Binary systems with such high redshifted total masses show fewer inspiral cycles in band prior to the merger, where large eccentricities seem astrophysically unlikely, since they would require interactions with third bodies at late times during the inspiral due to the rapid shedding of eccentricity 
\cite{PhysRev.136.B1224}.

To check whether these high eccentricity values come from non-physical artifacts associated with the short duration of the waveforms, we performed additional PE runs lowering both the reference frequency 
and the initial frequency of the waveform generator to 5~\si{Hz}. 
Furthermore, the \IMRPhenomTEHM model employs eccentricity-expanded \pn{} approximations for the description of the inspiral, which limit the applicability of the model up to eccentricities of
approximately $e\sim 0.4$ at an orbit-averaged GW frequency of 10~\si{Hz}~\cite{Planas:2025feq}.
For this reason, we have also compared our results with PE runs constraining the eccentricity prior up to $e_{\si{10Hz}}=0.3$.

\begin{figure}[t!]
    \centering
    \includegraphics[width=\columnwidth]{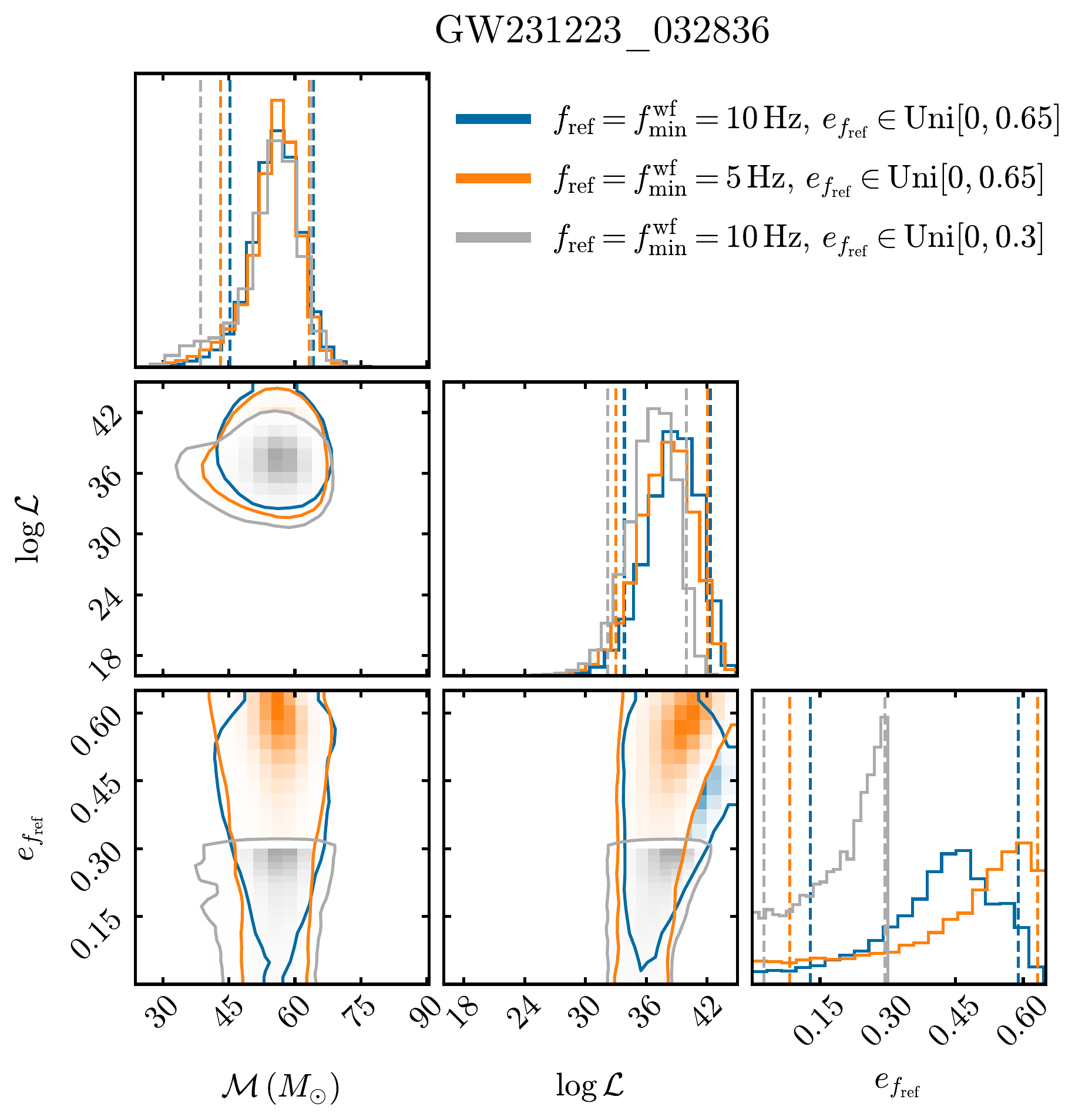}
    \caption{
     Comparison between some marginalized one- and two-dimensional posterior probability distributions obtained for GW231223\_032836 with different settings indicated with colors. 
     We show the posterior distributions for the redshifted chirp mass, $\mathcal{M}$, log likelihood, $\log\mathcal{L}$, and for the orbital eccentricity at the specified reference frequency, $e_{f_{\rm{ref}}}$.
     As in Fig.~\ref{fig:all_ecc_posteriors}, the eccentricity has been measured using the eccentric multipolar phenomenological \IMRPhenomTEHM model~\cite{Planas:2025feq}.
     The dashed vertical lines in the one-dimensional posteriors represent the $90\%$ credible intervals of each distribution. 
    }
    \label{fig:extra_ecc_runs_gw231223_032836}
\end{figure}

\begin{figure*}[t!]
    \centering
    \includegraphics[width=\textwidth]{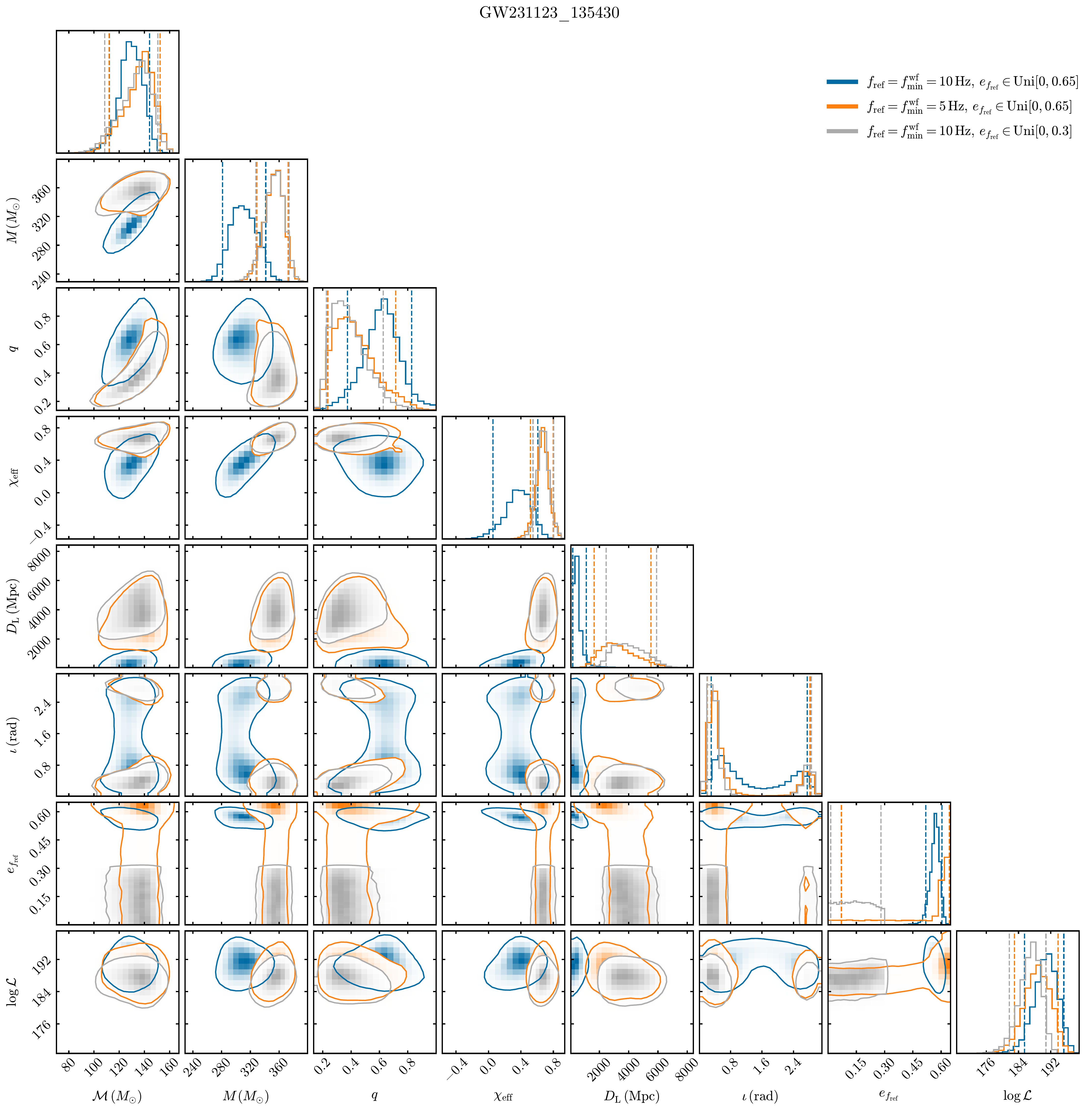}
    \caption{
     Comparison between the marginalized one- and two-dimensional posterior probability distributions obtained for GW231123\_135430 with different settings indicated with colors. 
     We show the posterior distributions for the redshifted chirp mass, $\mathcal{M}$, redshifted total mass, $M$, mass-ratio, $q$, effective inpsiral spin, $\chi_{\rm{eff}}$, luminosity distance, $D_{\rm{L}}$, inclination angle, $\iota$, orbital eccentricity at the specified reference frequency, $e_{f_{\rm{ref}}}$, and log likelihood, $\log\mathcal{L}$.
     As in Figs.~\ref{fig:all_ecc_posteriors} and~\ref{fig:extra_ecc_runs_gw231223_032836}, the eccentricity has been measured using the eccentric multipolar phenomenological \IMRPhenomTEHM model~\cite{Planas:2025feq}.
     The dashed vertical lines in the one-dimensional posteriors represent the $90\%$ credible intervals of each distribution. 
    }
    \label{fig:extra_ecc_runs_gw231123_135430}
\end{figure*}

As a representative case for these additional PE studies, we show 
the results for GW231223\_032836 in Fig.~\ref{fig:extra_ecc_runs_gw231223_032836}.
As seen from the marginal probability distributions, we infer larger eccentricity values when lowering $f_{\rm{ref}}$ and $f_{\rm{min}}^{\rm{wf}}$ from 10~\si{Hz} (orange posteriors) to 5~\si{Hz} (blue posteriors), while maintaining consistent 90\% credible intervals for $\mathcal{M}$ and $\log\mathcal{L}$ (and also for the rest of the parameters not shown).\footnote{In Figs.~\ref{fig:extra_ecc_runs_gw231223_032836} and~\ref{fig:extra_ecc_runs_gw231123_135430} we do not apply any of the postprocessing/reweighting techniques for the $\log\mathcal{L}$ described in App.~\ref{app:likelhood}. This is because we are not comparing with the GWTC-4.0 $\log\mathcal{L}$ values, but rather between several of our own PE runs, that share the same \BILBY version and sampling rate.} 
This behavior is also observed in GW231221\_135041 and suggests that the recovered eccentricity at $f_{\rm{ref}}=10\,\si{Hz}$ is unlikely to be an artifact caused by the short duration of the waveforms, since by extending the analysis to lower frequencies we allow for longer waveforms with more orbital cycles in the inference pipeline, which reduces potential artifacts originated by short-duration waveforms. 
Therefore, if such artifacts had dominated the sampling process, setting $f_{\rm{ref}}=f_{\rm{min}}^{\rm{wf}}= 5\,\si{Hz}$, would likely have caused the inferred eccentricity to vanish, which is not the case for the blue eccentricity posterior distribution shown in Fig.~\ref{fig:extra_ecc_runs_gw231223_032836}.
As an additional check, we show in gray the inferred posterior distributions with $f_{\rm{ref}}=f_{\rm{min}}^{\rm{wf}}= 10\,\si{Hz}$ and a reduced prior in eccentricity up to $e_{10\si{Hz}}=0.3$.
For this case, the $\log\mathcal{L}$ is shifted toward comparatively smaller values, and the eccentricity posterior rails toward the upper cut of the prior, which is consistent with hints of eccentricity greater than 0.3 at 10~\si{Hz}. 
These studies support the presence of eccentricity in GW231223\_032836 and GW231221\_135041. More detailed eccentric evolution consistency tests (see e.g.~\cite{Bhat:2025lri}) or single-detector re-analysis of the candidates are left for future work.

However, the results of the additional PE runs for the candidates GW231123\_135430 and GW230712\_090405
do not conclusively support a measure of eccentricity in the signal. 
As a representative example, Fig.~\ref{fig:extra_ecc_runs_gw231123_135430} shows the posterior distributions for GW231123\_135430, obtained using the same PE settings as in Fig.~\ref{fig:extra_ecc_runs_gw231223_032836}. 
While lowering $f_{\rm{ref}}$ from 10~\si{Hz} to 5~\si{Hz} leads to the recovery of higher eccentricity values, we observe substantial variations in the 90\% credible intervals for the rest of the parameters. A similar trend is seen for GW230712\_090405, where the results also depend sensitively on the chosen reference frequency.
This behavior may arise because the orbital eccentricity has not completely decayed by the time of merger, leading to inaccuracies in the \IMRPhenomTEHM model, which assumes QC mergers.  The understanding of the interplay between the data quality issues and the lack of eccentricity effects at merger-ringdown for high-mass events would require performing NR eccentric injections in real detector data and studying how the noise artifacts affect parameter degeneracies for eccentric signals and how the reference frequency choice influences the inferred posteriors (see~\cite{Miller:2025eak} for a detailed analysis in the context of precessing signals). We leave  such investigations for future work.  

\subsubsection{Eccentric candidates with medium values of eccentricity }\label{sec:results_medium_eccentricity:discussion}

The second group of eccentric candidates shown in Fig.~\ref{fig:all_ecc_posteriors} is formed by GW231224\_024321, GW230706\_104333, and GW231114\_043211 that show medium values of eccentricity $e_{10\si{Hz}}<0.3$.
These events correspond to comparatively lower redshifted total mass binaries, with inferred values below $45\,M_\odot$. 
The comparatively lower redshifted total mass results in a longer signal within the LVK sensitive band. This extended inspiral phase provides a better opportunity to constrain the orbital eccentricity, in contrast to high mass systems whose signals are dominated by the merger–ringdown, where state-of-the-art eccentric waveform models assume orbital circularization. Although these events are more likely to have residual orbital eccentricity by the time they enter the LVK band compared to the high-mass events described above, we do not have enough evidence to claim detection of eccentricity.

Here we also compute $e_{\rm gw}$ and $l_{\rm gw}$, obtained with \texttt{gw-eccentricity}~\cite{Shaikh:2023ypz}, for all the events reported in Fig.~\ref{fig:all_ecc_posteriors}. We find that the overall conclusions depicted by Fig.~\ref{fig:all_ecc_posteriors} do not vary substantially apart from the small expected shift due to a different eccentricity definition. 
For instance, for GW231226\_101520 the orbital eccentricity is $e_{10\si{Hz}} = 0.07^{+0.12}_{-0.01}$ and  $e_{\rm gw} = 0.06^{+0.06}_{-0.06}$. 
This agreement is also a consequence of the use of EOB coordinates in the definition of eccentricity in \IMRPhenomTEHM~\cite{Planas:2025feq}, which has been shown in Ref.~\cite{Gamboa:2024hli} to have good agreement with $e_{\rm{gw}}$. 
For the high redshifted mass eccentric candidates presented above, the calculation of GW eccentricity is challenging due to the short duration of the inspiral and the higher eccentricity of the waveforms. 
One approach for the calculation of GW eccentricity on a posterior sample would be to evaluate the \IMRPhenomTEHM model for the same binary parameters keeping the reference frequency, but lowering the starting frequency of waveform generation. However, for the eccentric \IMRPhenomTEHM model this implies going to increasingly high eccentricity, where the missing eccentric harmonics in the waveform (the model is expanded up to $\mathcal{O}(e^6)$) introduce some interference pattern due to the missing eccentric harmonics which complicates the calculation of GW eccentricity. Thus, we leave for the future the improvement of the eccentric inspiral description of the \IMRPhenomTEHM model by implementing non-eccentricity expanded expressions for which GW eccentricity can be computed for high eccentricities.

\section{Conclusions}\label{sec:conclusions}

\begin{table*}[htbp]
\caption{Summary of the relevant information and parameters for all the events discussed in detail in this work. In the data quality (DQ) issues column, (G) indicates glitch mitigation applied, while (B) indicates bandwidth reduction. In the systematics column, we indicate with~$\checkmark$ if the event displays waveform systematic differences in any result reported in the \GWTCFOUR analysis or in our analysis, and with (M) if the event presents also multimodality for some of the inferred quantities. In the eccentric–type (ecc. type) column, (H) labels high–total–mass cases and (L) denotes comparatively lower–total–mass cases as discussed in Sec.~\ref{sec:results_eccentricity:discussion}.}
\label{tab:ordered}
\begin{ruledtabular}
\begin{tabular}{lcccccccccc}
Event & \textbf{DQ issues} & systematics & SNR H1 & SNR L1 & $M\,(M_{\odot})$ & $\chi_{\rm{eff}}$ \
& $\chi_{\rm{p}}$ & \textbf{ecc. type} & $\mathcal{B}_{\log10}^{\rm{TEHM/THM}}$ & $\mathcal{B}_{\log10}^{\rm{TEHM/TPHM}}$ \\
\hline
GW230624\_113103 &  & $\checkmark$ & \ensuremath{7.1^{+1.4}_{-1.5}} & \
\ensuremath{6^{+1.2}_{-1.2}} & \ensuremath{57.6^{+8.9}_{-5}} & \
\ensuremath{0.1^{+0.3}_{-0.2}} & \ensuremath{0.4^{+0.4}_{-0.3}} &  &  \
&  \\
GW230628\_231200 &  & $\checkmark$ & \ensuremath{11.5^{+1.4}_{-1.3}} & \
\ensuremath{9.9^{+1.2}_{-1.1}} & \ensuremath{82.2^{+6}_{-5.7}} & \
\ensuremath{0^{+0.1}_{-0.2}} & \ensuremath{0.5^{+0.4}_{-0.4}} & \
 &  &  \\
GW230704\_212616 &  & $\checkmark$ & \ensuremath{5.3^{+1.3}_{-1.3}} & \
\ensuremath{5.4^{+1.4}_{-1.3}} & \ensuremath{286^{+44}_{-51}} & \
\ensuremath{0.3^{+0.3}_{-0.4}} & \ensuremath{0.4^{+0.4}_{-0.3}} &  &  \
&  \\
GW230706\_104333 &  &  & \ensuremath{5.1^{+2}_{-1.2}} & \
\ensuremath{6.8^{+1.7}_{-1.7}} & \ensuremath{37.4^{+3.1}_{-1.5}} & \
\ensuremath{0.2^{+0.1}_{-0.2}} & \ensuremath{0.5^{+0.4}_{-0.3}} & \
(L) & \ensuremath{0.23^{+0.14}_{-0.14}} & \
\ensuremath{-0.04^{+0.14}_{-0.14}} \\
GW230712\_090405 &  & (M) & \ensuremath{4.8^{+1.3}_{-1.2}} & \
\ensuremath{6.1^{+1.6}_{-1.5}} & \ensuremath{67^{+35}_{-18}} & \
\ensuremath{-0.1^{+0.4}_{-0.3}} & \ensuremath{0.6^{+0.3}_{-0.4}} & (H) & \
\ensuremath{0.76^{+0.14}_{-0.14}} & \ensuremath{0.09^{+0.14}_{-0.14}} \
\\
GW230723\_101834 &  & (M) & \ensuremath{6.9^{+1.4}_{-1.7}} & \
\ensuremath{6.1^{+1.5}_{-1.2}} & \ensuremath{35.2^{+4}_{-1.7}} & \
\ensuremath{-0.2^{+0.2}_{-0.2}} & \ensuremath{0.4^{+0.4}_{-0.3}} &  & \
 &  \\
GW230814\_230901 &  & $\checkmark$ & 0 & \ensuremath{41.7^{+1.6}_{-1.7}} & \
\ensuremath{65.6^{+1.7}_{-1.9}} & \ensuremath{0^{+0}_{-0.1}} & \
\ensuremath{0.3^{+0.3}_{-0.2}} &  &  &  \\
GW230927\_153832 &  & $\checkmark$ & \ensuremath{12.4^{+1.2}_{-1.2}} & \
\ensuremath{14.8^{+1.5}_{-1.3}} & \ensuremath{46.9^{+1.7}_{-1.2}} & \
\ensuremath{0^{+0.1}_{-0.1}} & \ensuremath{0.4^{+0.4}_{-0.3}} &  &  & \
 \\
GW231028\_153006 &  & $\checkmark$ (M) & \ensuremath{11.2^{+1.1}_{-1.2}} & \
\ensuremath{17.4^{+1.4}_{-1.5}} & \ensuremath{252^{+19}_{-25}} & \
\ensuremath{0.4^{+0.2}_{-0.2}} & \ensuremath{0.5^{+0.3}_{-0.3}} & \
 &  &  \\
GW231114\_043211 & (G) &  & \ensuremath{6.6^{+1.6}_{-1.5}} & \
\ensuremath{6.4^{+1.6}_{-1.3}} & \ensuremath{38.7^{+9.6}_{-4}} & \
\ensuremath{0.1^{+0.2}_{-0.2}} & \ensuremath{0.2^{+0.3}_{-0.2}} & (L) & \
\ensuremath{0.22^{+0.15}_{-0.15}} & \ensuremath{0.66^{+0.15}_{-0.15}} \
\\
GW231118\_005626 & (B) & $\checkmark$ & \ensuremath{7.1^{+1.3}_{-1.3}} & \
\ensuremath{7.3^{+1.3}_{-1.4}} & \ensuremath{42.5^{+4.1}_{-2.3}} & \
\ensuremath{0.4^{+0.1}_{-0.2}} & \ensuremath{0.5^{+0.3}_{-0.3}} &  &  \
&  \\
GW231118\_090602 & (G) & $\checkmark$ (M) & \
\ensuremath{7.2^{+1.3}_{-1.2}} & \ensuremath{7.6^{+1.3}_{-1.3}} & \
\ensuremath{25.6^{+10}_{-1.5}} & \ensuremath{0.1^{+0.3}_{-0.1}} & \
\ensuremath{0.4^{+0.4}_{-0.3}} &  &  &  \\
GW231123\_135430 & (G) & $\checkmark$ (M) & \
\ensuremath{12.6^{+1.6}_{-1.3}} & \ensuremath{15.7^{+1.5}_{-1.6}} & \
\ensuremath{292^{+37}_{-24}} & \ensuremath{0.2^{+0.3}_{-0.3}} & \
\ensuremath{0.8^{+0.1}_{-0.3}} & (H) & \ensuremath{2.6^{+0.15}_{-0.15}} \
& \ensuremath{-4.92^{+0.16}_{-0.16}} \\
GW231221\_135041 & (G) &  & \ensuremath{5.8^{+1.5}_{-1.5}} & \
\ensuremath{4.5^{+1.3}_{-1.1}} & \ensuremath{131^{+18}_{-18}} & \
\ensuremath{0^{+0.4}_{-0.3}} & \ensuremath{0.6^{+0.3}_{-0.4}} & \
(H) & \ensuremath{0.36^{+0.12}_{-0.12}} & \
\ensuremath{-0.07^{+0.12}_{-0.12}} \\
GW231223\_032836 & (G) &  & \ensuremath{5.8^{+1.6}_{-1.3}} & \
\ensuremath{6^{+1.6}_{-1.4}} & \ensuremath{130^{+19}_{-22}} & \
\ensuremath{-0.2^{+0.3}_{-0.3}} & \ensuremath{0.5^{+0.3}_{-0.4}} & \
(H) & \ensuremath{0.57^{+0.13}_{-0.13}} & \
\ensuremath{0.53^{+0.13}_{-0.13}} \\
GW231224\_024321 & (B) &  & \ensuremath{8.3^{+1.2}_{-1.2}} & \
\ensuremath{9.3^{+1.3}_{-1.4}} & \ensuremath{19.6^{+1}_{-0.2}} & \
\ensuremath{0^{+0.1}_{-0.1}} & \ensuremath{0.3^{+0.5}_{-0.2}} & \
(L) & \ensuremath{0.30^{+0.15}_{-0.15}} & \ensuremath{0.43^{+0.16}_{-0.16}} \\
GW231226\_101520 &  & $\checkmark$ & \ensuremath{25.5^{+1.3}_{-1.3}} & \
\ensuremath{21.4^{+1.1}_{-1.1}} & \ensuremath{91^{+3.6}_{-3.4}} & \
\ensuremath{-0.1^{+0.1}_{-0.1}} & \ensuremath{0.4^{+0.3}_{-0.3}} & \
 &  &  \\
\end{tabular}
\end{ruledtabular}
\end{table*}

In this work, we have extended the parameter estimation analysis of the \GWTCFOUR catalog by performing systematic parameter estimation with four state-of-the-art phenomenological waveform models: 
\IMRPhenomTEHM models aligned-spin eccentric inspirals, and allows us to identify candidates for eccentricity;
\IMRPhenomTPHM is a QC precessing model that agrees with \IMRPhenomTEHM in the aligned spin QC limit, described by 
\IMRPhenomTHM. Combining results from these three models allows us to compute model selection Bayes factors between consistent models that describe either spin precession or eccentricity. 
The fourth model, \IMRPhenomXPNR provides an improved description of spin precession in the frequency domain, which is calibrated to \nr{} simulations
with misaligned spins.

Using these models, we re-analyzed all 84 BBH events in \GWTCFOUR{}, producing a homogeneous catalog of posterior samples that complements the LVK results and enables consistent Bayes-factor comparisons between eccentric and precessing hypotheses.

For QC waveform models, we find excellent agreement among \IMRPhenomXPNR{}, \IMRPhenomTPHM{}, and the \GWTCFOUR{} posteriors for the large majority of events.
Across our three summary metrics—the Jensen–Shannon divergence, median-shift statistic, and credible-interval overlap—only a small subset of sources exhibit waveform-model differences exceeding statistical uncertainties.
These outliers are predominantly high–total-mass systems with limited inspiral content, where the measured SNR is dominated by the merger–ringdown and where current waveform models rely most heavily on extrapolation beyond the NR calibration region.
In these cases, we observe measurable shifts in the inferred masses and spins between different precessing models.
Consistent with the findings of the LVK, \IMRPhenomXPHMST{} tends to recover lower likelihoods for such sources, while the newer models—particularly \IMRPhenomXPNR{} and \IMRPhenomTPHM{}—yield improved  consistency in the likelihood distribution, suggesting they provide a more reliable description of strong-precession and high-mass systems than \IMRPhenomXPHMST{}. Nevertheless, results from \IMRPhenomXPNR{} and \IMRPhenomTPHM{} still display systematic differences for the most massive and spinning events in several parameters, in agreement with the systematic differences observed between models in \GWTCFOUR{}. Furthermore, \IMRPhenomTPHM{} show additional systematic differences for two events, and we traced the reason for one of them (GW230814\_230901, a loud single-detector signal observed with edge-on inclination) to be the missing $(3,2)$ harmonic content in the model, highlighting the need to further complete the physical content of the models.

Most \bbh{} events in \GWTCFOUR{} do not show evidence for eccentricity: their \IMRPhenomTEHM{} posteriors are consistent with $e=0$ measured at a reference frequency of 10~\si{Hz}, and the Bayes factors disfavour eccentricity relative to the QC models.
A small number of systems, listed in Table~\ref{tab:ordered}, warranted additional scrutiny either because they show mild support for eccentricity or because they exhibit waveform-model systematics.
We have identified a total of 7 eccentric candidates that produce strictly positive Bayes factors for eccentricity when compared to the aligned-spin QC model. We have divided them into two groups: GW231123\_135430, GW230712\_090405, GW231223\_032836, and GW231221\_135041, which have an inferred redshifted total mass above $90\,M_{\odot}$ and a moderately large $e_{10\si{Hz}}>0.4$; and GW231224\_024321, GW230706\_104333, and GW231114\_043211, which show a comparatively lower redshifted total mass $M<45\,M_{\odot}$ and lower eccentricity $e_{10\si{Hz}}<0.3$.
For the majority of these events, the preference of eccentricity is drastically reduced once the comparison is made against the QC precessing hypothesis, which confirms the expected degeneracy between moderate eccentricity and precessional dynamics at low \pn{} orders.
Only three of these sources—GW231223\_032836, GW231114\_043211, and GW231224\_024321—continue to show significant support for eccentricity even relative to precessing QC models. 
However, the interpretation of these results is complicated by the proximity of the reference frequency (at which the eccentricity is measured) to a band of glitch-subtracted data for GW231223\_032836 and GW231114\_043211, and by the removal of contaminated data below 40~\si{Hz} in LIGO-Hanford at the time of GW231224\_024321.

Our analysis clarifies several aspects of waveform systematics.
First, discrepancies between models arise primarily in regimes with limited inspiral cycles or high spins.
Second, we find good agreement between frequency-domain and time-domain precessing models across most of the catalog, indicating that the dominant uncertainties are physical—related to high-spin and high-mass systems—rather than arising from the choice of modeling domain.
Third, for aligned-spin analyses, \IMRPhenomTEHM{} provides a robust description of eccentric inspirals, but distinguishing genuine eccentricity from precessional covariances requires waveform models that combine both effects consistently.

Taken together, our results show that (i) eccentricity remains largely unconstrained for most O4 BBH events; (ii) claims of moderate eccentricity are highly covariant with precession unless data quality and waveform systematics are exceptionally well controlled; and (iii) joint precessing–eccentric models with higher multipoles and extended NR calibration will be essential for unambiguous identification of dynamical-capture binaries in future observing runs.

All posterior samples, configuration files, and analysis scripts have been made publicly available on Zenodo.
Our automated framework \PEAUTOMATOR{} enables reproducible large-scale parameter-estimation campaigns across heterogeneous HPC resources.
The extended catalog presented here provides a consistent data set for future studies of waveform systematics, population inference, and astrophysical modeling with the \GWTCFOUR{} events.

\section*{Author contributions}
SH, YX, ARB contributed to the formulation or evolution of overarching research goals and aims. 
YX, SH, ARB contributed to the management and coordination responsibility for the research activity planning and execution.
YX designed and implemented the software for automatization of simulations. 
MRS implemented the automatization of result visualization. 
MdLP and YX contributed to the environment setup on clusters.
SH contributed the original drafts of Secs.~\ref{sec:Introduction}, \ref{sec:conclusions}. 
MC, EH and ARB provided the original draft of Sec.~\ref{sec:models}.
YX and SH provided the original draft of Sec.~\ref{sec:methods}.
HE led the formal analysis, visualization, and original draft for section~\ref{sec:results:QC}. MRS contributed to the formal analysis and visualization for section~\ref{sec:results:QC}.
ARB and JV led the formal analysis, visualization, and original draft for section~\ref{sec:results_eccentricity}. YX contributed to App.~\ref{app:likelhood}.
ARB, EH, MC, JLQ, JV, MRS, SH, YX, HE contributed to the review and editing of the paper.
MC, YX, SH, EH contributed to the acquisition of computing resources.
ARB, EH, JV, MdLP, MRS, SH contributed to waveform model validation and methodology drafting.

All AUTHORS contributed to the data analysis with more than a thousand parameter estimation runs and valuable discussion in around 30 meetings.

\section*{Acknowledgements}

The authors would like to thank the reviewer Shubhanshu Tiwari for the LSC Publication \& Presentation Committee review of this manuscript. We thank the useful discussion with Charlie Hoy on calibration envelope. We thank Cecilio García Quirós for discussion on the phenomxpy code. This paper has document number
LIGO-P2500737.

We thankfully acknowledge the computer resources (MN5 Supercomputer), technical expertise and assistance provided by Barcelona Supercomputing Center (BSC)  through funding from the Red Española de Supercomputación (RES) (AECT-2025-2-0045); and the computer resources (Picasso Supercomputer), technical expertise and assistance provided by the SCBI (Supercomputing and Bioinformatics) center of the University of Málaga (AECT-2025-2-0025, AECT-2025-2-0029).
This research has made use of data or software obtained from the Gravitational Wave Open Science Center (gwosc.org), a service of the LIGO Scientific Collaboration, the Virgo Collaboration, and KAGRA.
This material is based upon work supported by NSF's LIGO Laboratory which is a major facility fully funded by the National Science Foundation.
LIGO is funded by the U.S. National Science Foundation. Virgo is funded by the French Centre National de Recherche Scientifique (CNRS), the Italian Istituto Nazionale della Fisica Nucleare (INFN) and the Dutch Nikhef, with contributions by Polish and Hungarian institutes.

This work was supported by the Universitat de les Illes Balears (UIB); the Spanish Agencia Estatal de Investigación grants PID2022-138626NB-I00, PID2019-106416GB-I00, RED2022-134204-E, RED2022-134411-T, funded by MCIN/AEI/10.13039/501100011033; the MCIN with funding from the European Union NextGenerationEU/PRTR (PRTR-C17.I1); Comunitat Autonòma de les Illes Balears through the Direcció General de Recerca, Innovació I Transformació Digital with funds from the Tourist Stay Tax Law (PDR2020/11 - ITS2017-006), the Conselleria d’Economia, Hisenda i Innovació grant numbers SINCO2022/18146 and SINCO2022/6719, co-financed by the European Union and FEDER Operational Program 2021-2027 of the Balearic Islands; the “ERDF A way of making Europe”. 

A. Ramos-Buades was supported by the Veni research programme which is (partly) financed by the Dutch Research Council (NWO) under the grant VI.Veni.222.396,  PID2024-157460NA-I00 funded by MICIU/AEI/10.13039/501100011033 and the ERDF/EU; and the Spanish Ministerio de Ciencia, Innovación y Universidades (Beatriz Galindo, BG23/00056), co-financed by UIB.
J. Valencia, M. Rosselló-Sastre and
M. de Lluc Planas
were supported by the Spanish Ministry of Universities, grants FPU22/02211, FPU21/05009,
and FPU20/05577. 
F. A. Ramis Vidal and J. Llobera Querol were supported by the Conselleria d'Educació i Universitats del Govern de les Illes Balears via FPI-CAIB doctoral grants FPI\_092\_2022 and FPI\_093\_2022.
H. Estellés was supported by the Spanish Ministerio de Ciencia, Innovación y Universidades (Juan de la Cierva, JDC2023-051934-I).
E. Hamilton and M. Colleoni were financed by the Conselleria d’Educació i Universitats del Govern de les Illes Balears and the European Social Fund Plus (ESF+) 2021-2027, grants
POSTDOC2024\_25 and 
POSTDOC2024\_52.
A. Heffernan was supported by grant PD-034-2023 co-financed by the Govern Balear and the European Social Fund Plus (ESF+) 2021-2027.
A. Montava was funded by Programa SOIB-INVESTIGO (RI-24) Govern de les Illes Balears with  funds from the European Union - NextGenerationEU/PRTR-C23.I01.P03.S02. 
M. de Lluc Planas's research was supported by the European Research Council (ERC) Horizon Synergy Grant “Making Sense of the Unexpected in the Gravitational-Wave Sky” grant agreement no. GWSky–101167314.

\appendix

\section{Likelihood re-weight}\label{app:likelhood}

A consistent definition of the likelihood is required when comparing results across different parameter–estimation runs, particularly when interpreting likelihood values or Bayes factors. In our case, two independent issues necessitate a reweighting of the likelihoods: (i) the likelihood normalization error present in the \GWTCFOUR{} posterior samples, and (ii) a discrepancy introduced by differing data–resampling algorithms between the \GWTCFOUR{} analysis and the PE runs performed in this work.

The reweighting procedure consists of two components. First, this work employs the corrected version of \BILBY{}, whereas the \GWTCFOUR{} posterior samples were produced with an earlier \BILBY{} version affected by an incorrect likelihood normalization.
As described in \cite{LIGOScientific:2025yae}, this error can be corrected by rescaling the likelihood according to
\begin{equation}
\hat{p}(d \mid \boldsymbol{\theta}) = p(d \mid \boldsymbol{\theta})^{\beta},
\end{equation}
where
\begin{equation}\label{eq:beta_factor}
\beta = \left( 1 - \frac{5 T_w}{4 T} \right)^{-1},
\end{equation}
with $T_w$ denoting the Tukey window duration and T the total segment length. This correction ensures consistency between our likelihood evaluations and those underlying the \GWTCFOUR{} samples. 
The \GWTCFOUR{} Zenodo release \cite{ligo_scientific_collaboration_and_virgo_2025_17014085} currently provides two versions: the initial release (v1), generated with the uncorrected \BILBY{} likelihood implementation, and a subsequent release (v2), in which the posterior samples have been reweighted to incorporate the corrected likelihood normalization described above.

The second component addresses a mismatch between likelihood evaluations obtained using 4 kHz data (used in this work) and those based on 16 kHz data (used in the \GWTCFOUR{} runs). The discrepancy, which propagates to the Bayesian evidence and Bayes factors, originates from the distinct resampling procedures: \BILBY{} employs the \LALSUITE{} resampling, while the GWOSC-provided downsampled strain data use \texttt{scipy.signal.decimate}, which applies an internal anti-aliasing filter. This filtering affects likelihood calculations but leaves the posterior distributions largely unchanged.

\begin{figure}[t!]
    \vspace{3mm}
    \centering
    \includegraphics[width=\linewidth]{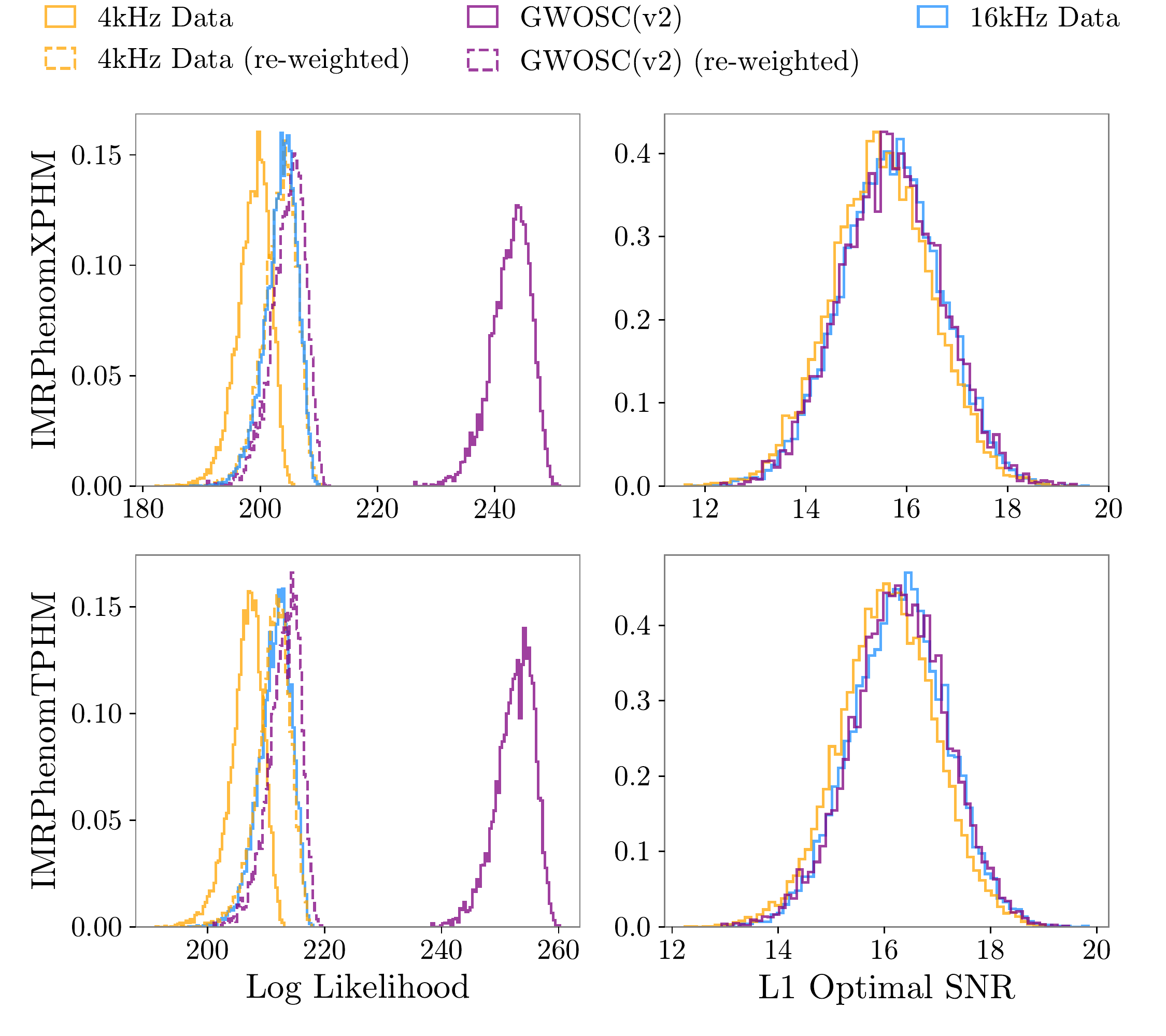}
    \caption{Comparison between analyses performed with GWOSC 4 kHz data, GWOSC 16 kHz data, and the GWOSC-provided posterior samples for GW231123\_135430. Results are shown for two waveform models, \IMRPhenomXPHM{} and \IMRPhenomTPHM{}. The left panels display the log-likelihood distributions, while the right panels show the L1 optimal SNR. The yellow histograms correspond to runs using the 4 kHz data; the solid yellow histograms denote the original likelihood evaluations, and the dashed yellow histograms show the corresponding values after recomputing the likelihood with the 16 kHz data. The blue histograms represent independent reruns performed directly with 16 kHz data. The purple histograms show the GWOSC(v2) posterior samples, with the dashed purple histograms indicating the likelihoods after applying the $\beta$-reweighting described in Eq. \eqref{eq:beta_factor}.
    }
    \label{fig:likelihood-reweight}
\end{figure}

To reconcile the two settings without rerunning all parameter–estimation jobs at 16 kHz—an approach that would be computationally prohibitive—we recompute the likelihoods for the existing posterior samples using \BILBY{}’s built-in likelihood evaluation. We performed targeted verification runs using 16 kHz GWOSC data and confirmed that the reweighted 4 kHz likelihoods match the corresponding 16 kHz results to numerical precision. An illustrative comparison is shown in Fig.~\ref{fig:likelihood-reweight}. The residual discrepancy between the reweighted GWOSC results and our 16 kHz reruns likely arises from the likelihood–normalization error present in the earlier \BILBY{} release used to generate the GWOSC samples and statistical error.

\let\c\Originalcdefinition %
\let\d\Originalddefinition %
\let\i\Originalidefinition

\bibliography{bib}

\end{document}